\documentclass{pasj00}
\draft

\begin{document}
\SetRunningHead{Author(s) in page-head}{Running Head}

\title{Modeling the spectral-energy-distribution of 3C 454.3 in a ``flat" broad-line-region scenario}


\author{Maichang Lei}
\affil{Yunnan Observatory, Chinese Academy of Sciences,  Kunming 650011, China}
\email{maichanglei83@ynao.ac.cn}

\author{Jiancheng Wang}
\affil{Yunnan Observatory, Chinese Academy of Sciences,  Kunming 650011, China}
\email{jcwang@ynao.ac.cn}

%


\maketitle

\begin{abstract}
The broad-line region (BLR) of flat-spectrum radio quasars (FSRQs) could have a ``flat" geometrical structure to allow GeV gamma-ray photons escape, to produce the observed gamma-ray flares with short timescales. In this paper, we collect the quasi-simultaneous spectral energy
distributions (SEDs) of the FSRQ 3C 454.3 obtained by the multi-wavelength campaigns spanning from 2007 July to 2011 January, and use a model with the ``flat" structure BLR, the accretion disc and  the dust torus to explain the SEDs of gamma-ray outbursts. We obtain the following results:
(i) The jet is almost in equipartition between magnetic and particle energy densities during the outbursts;
(ii) When the emitting region locates inside the cavity of the BLR, the covering factor $f_{\rm BLR}$ of the BLR is very small; as the emitting region goes into the BLR structure, $f_{\rm BLR}$ increases.
(iii) The aperture angle $\alpha$ describing the BLR structure is about $45^{\circ}$;
(iv) The central black hole (BH) mass is about $5\times 10^{8}$\,$M_{\odot}$ rather than $4.4\times 10^{9}$\,$M_{\odot}$.
\end{abstract}

\section{Introduction}
Blazars constitute a subclass of radio-loud active galactic nuclei (AGNs) characterized by broadband non-thermal emission from radio to $\gamma$-rays, which is widely considered to be produced by bipolar, relativistic jets closely aligned to our line of sight. They exhibit rapid variability, high degree of polarization, and two distinct spectral humps in the $\nu-\nu F_{\nu}$ representation. Blazars with only weak or entirely absent emission lines in the optical band are classified as BL Lacertae objects (BL Lacs), while otherwise are classified as flat-spectrum radio quasars (FSRQs)\citep{1995PASP..107..803U}.

FSRQ 3C 454.3 (PKS 2251+158; z=0.859: \cite{1991MNRAS.250..414J}) is one of the most active and brightest sources in the $\gamma$-ray sky. In 2005 April-May, it underwent a dramatic optical outburst, reaching R=12.0 \citep{2006A&A...453..817V}. Unfortunately, due to the lack of contemporaneous $\gamma$-ray data, no firm conclusions could be drawn on physical mechanisms responsible for the acceleration of radiating particles in the emitting region, the origin of the high-energy component, the jet power and the bolometric luminosity during the outbursts. Following this flare, 3C 454.3 has been the object of several multi-wavelength campaigns to follow the source emission behavior \citep{2006A&A...453..817V,2007A&A...464L...5V,2007A&A...473..819R,2008A&A...485L..17R,2008A&A...491..755R}. In 2007 July, 3C 454.3 exhibited the brightest $\gamma$-ray flare at that time \citep{2008ApJ...676L..13V}. During 2007, November and December, AGILE detected high $\gamma$-ray activity from this source, and triggered the corresponding multi-wavelength campaigns \citep{2009ApJ...690.1018V,2009ApJ...707.1115D,2009A&A...498...83A}. Subsequently, \cite{2010ApJ...712..405V} presented 18 months' multi-wavelength observations carried out from 2007 July to 2009 January. In particular, they showed the results of the AGILE campaigns on 2008 May-June, 2008 July-August, and 2008 October-2009 January. In the first week of 2009 December, 3C 454.3 became the brightest $\gamma$-ray source in the sky with  $10^{-5}$\,photons~cm$^{-2}$~s$^{-1}$ above 100\,MeV. The corresponding multifrequency campaign results were reported by \cite{2010ApJ...716L.170P} and \cite{2011MNRAS.410..368B} with different time span.

Since the launch of the Large Area Telescope (LAT) on board the Fermi Gamma-ray Space Telescope \citep{2009ApJ...697.1071A}, 3C 454.3 has been continuously monitored in the GeV energy band. The Fermi-LAT observation has conclusively confirmed that the GeV spectrum in the multifrequency campaign on 2008 August did not follow a power-law and instead showed a clear break at energy $E_{\rm br}=2.4\pm 0.3$\,GeV \citep{2009ApJ...699..817A}. \cite{2010ApJ...714L.303F} showed that such break could be satisfactorily explained as a combination of Compton-scattered disc and BLR radiations. \cite{2010ApJ...717L.118P} and \cite{2011MNRAS.417L..11S} attributed this break to photon-photon absorption produced by He II and H I Lyman recombination continua (LyC), this result limits the $\gamma$-ray emission region close to the boundary of the high-ionization part of the BLR. \cite{2012ApJ...761....2H} showed that the scenario proposed by \cite{2010ApJ...717L.118P} is not really consistent with the Fermi observation. Observations performed with the Fermi-LAT telescope have revealed the presence of spectral break in the GeV spectrum of FSRQs and LBLs and IBLs. \cite{2013ApJ...771L...4C} proposed that this feature could be explained by Compton scattering of BLR photons by relativistic electrons with a log-parabolic distribution.

The BLR around the central engine of FSRQs produces the diffuse soft photons with characteristic frequency peaking at $\sim 2.47\times 10^{15}$\,Hz via reverberation mechanism. Those soft photons can be scattered up to GeV energies by relativistic electrons within the jet. At the same time, those soft photons also absorb the $\gamma$-ray photons from the emission region via the photon-annihilation/pair-creation process $\gamma+\gamma \to e^{+}+e^{-}$. The absorption of GeV photons by the diffuse photon field of the BLR has been investigated by many authors \citep{2003APh....18..377D,2009MNRAS.392L..40T,2012arXiv1209.2291T}. The results showed that the photons from the BLR will significantly absorb the $\gamma$-rays with energy above a few tens of GeV. \cite{2006ApJ...653.1089L} used a spherical BLR with half-thickness $h$ (0.05$\to$ 0.3\,pc) to investigate the absorption of $\gamma$-rays, and found that the $\gamma$-rays with energies from 10 up to 200\,GeV cannot escape from the diffuse photon field. Recently, \cite{2014PASJ...66....7L} (hereafter, Lei14) generalized the spherical BLR structure to a ``flat" structure measured by an aperture angle $\alpha$, and found that the $\gamma$-rays with above mentioned energies could escape transparently even if the $\gamma$-ray emission region is located inside the cavity of the BLR. In this paper, we try to model the quasi-simultaneous SEDs of 3C 454.3 during the outbursts based on the Synchrotron-Self-Compton (SSC) plus External-Compton (EC) model, and constrain the BLR structure.

In Section 2, we give a brief description of the model. In Section 3, we apply the model fitting the SEDs of 3C 454.3 to obtain the BLR structure and the properties of the jet. Our discussion is given in Section 4. The conclusion and summary are given in Section 5. Symbols with a numerical subscript should be read as a dimensionless number $X_{\rm n}=X/(10^{\rm n}\,\rm cgs~units)$.
We adopt the cosmological parameters from the Planck Collaboration \citep{2013arXiv1303.5706A}: $H_{0}=67.3$\,km~s$^{-1}$~Mpc$^{-1}$, $\Omega_{\rm m}=0.315$, $\Omega_{\Lambda}=0.685$.

\section{The model}
\subsection{Synchrotron-Self-Compton emission}
The leptonic model has been successfully used to explain the multi-band SEDs of blazars. We assume
a spherical blob in the jet moving with bulk Lorentz factor $\Gamma_{\rm j}$ at a small angle $\theta_{\rm v}$ with the line of sight. The blob is filled with relativistic electrons (hereafter by electrons we mean both electrons and positrons) and a tangled, uniform magnetic field B. The observed radiation is strongly boosted by a Doppler factor $\delta_{\rm D}=[\Gamma_{\rm j}(1-\beta_{\Gamma}\cos\theta_{\rm v})]^{-1}$, where $\beta_{\Gamma}$ is the bulk velocity of the plasma in terms of the speed of light. The comoving size $R_{\rm b}$ of the blob is assumed to be determined by the observed variability timescale $t_{\rm var}$ as $R_{\rm b}\simeq t_{\rm var} {c\delta_{\rm D}}/{(1+z)}$. Throughout the paper, unprimed quantities refer to the observer's frame and primed quantities refer to the blob's frame, while the starred quantities refer to the stationary frame with respect to the BH.

We adopt a broken power-law function with a cutoff to describe the electron energy distribution in the emission region:
\begin{eqnarray}
n_{\rm e}(\gamma^{\prime}) &=& \frac{n_{0}{\gamma_{\rm b}^{\prime}}^{-1}}{(\gamma^{\prime}/\gamma_{\rm b}^{\prime})^{s_{1}}+(\gamma^{\prime}/\gamma_{\rm b}^{\prime})^{s_{2}}},
\end{eqnarray}
where $n_{0}$ is the number density of the electrons at $\gamma^{\prime}={1}$; $\gamma^{\prime}=E_{\rm e}^{\prime}/m_{\rm e}c^{2}$ is the electron Lorentz factor assumed to vary between $\gamma_{\rm min}^{\prime}\leq \gamma^{\prime} \leq \gamma_{\rm max}^{\prime}$. $s_{1}$ and $s_{2}$ are the electron power-law indexes below and above the break energy $\gamma_{\rm b}^{\prime}$, respectively.

Having the electron energy distribution, we use the formula given by \cite{2008ApJ...686..181F} and \cite{2009ApJ...692...32D} to calculate the observed synchrotron and SSC flux.

\subsection{External-Compton emission}
The external photons come from the BLR, the accretion disc and the dust torus. First, we calculate the photon field of the BLR. We assume that the BLR has a flat structure described by an angle $\alpha$ (measured from the disc plane), named as ``aperture angle" in Lei14. The energy density is thus given by
\begin{eqnarray}
U(R) &=& \frac{2\pi}{c}\int_{\theta_{\star,\rm min}}^{\theta_{\star,\rm max}} I(R,\theta_{\star})\sin\theta_{\star} d\theta_{\star},
\end{eqnarray}
where $I(R,\theta_{\star})$ is the radiation intensity of the BLR at angle $\theta$ to the jet axis and at position $R$, which is in units of $\rm erg~cm^{-2}~s^{-1}$; $\theta_{\star,\rm min}$ and ${\theta_{\star,\rm max}}$ are the minimum and maximum
angles seen by the blob (see Lei14 in detail).

Then, we calculate the photon field produced by the accretion disc. We assume the disc to be geometrically thin and optically thick. The energy density is
thus given by
\begin{eqnarray}
u_{\star}(\epsilon_{\star},\Omega_{\star}) &=& \frac{4}{2.7^{4}}~\frac{3}{32}\frac{R_{\rm s}L_{\rm d}}{\pi c \eta \alpha_{\rm r}} \frac{\varphi(r)}{r^{3}T_{\rm D}^{4}(r)}
\nonumber\\
&\times& \frac{2(m_{\rm e}c^{2})}{\lambda_{\rm c}^{3}}\frac{\epsilon_{\star}^{3}}{\exp(\epsilon_{\star}/\Theta)-1},
\end{eqnarray}
where $\alpha_{\rm r}\simeq 7.566\times 10^{-15}$\,erg cm$^{-3}$ K$^{-4}$ is the Radiation constant, $R_{\rm s}=2GM_{\rm BH}/c^{2}$ is the Schwarzschild radius;
$\eta=1/12$ is the efficiency to transform accreted matter to escaping radiation energy; $\varphi (r) \simeq 1-(3R_{\rm s}/r)^{1/2}$. $\Theta=2.7k_{\rm B}T_{\rm D}(r)/m_{\rm e}c^{2}$ is the dimensionless temperature of disc,
$T_{\rm D}(r)$ is the radial surface temperature profile \citep{1973A&A....24..337S,2009MNRAS.397..985G}.

Finally, we calculate the photon field produced by the dust torus. We assume that the dust torus has a spherical and thin structure with typical frequency at $\nu_{\star,\rm IR}=3\times 10^{13}$\,Hz, corresponding to the temperature as $T_{\rm IR}=h\nu_{\star,\rm IR}/(3.93k_{\rm B})$. The size is determined based on the formula given by \cite{2012MNRAS.420..526M}
\begin{eqnarray}
R_{\rm IR} &=& 1.6\times 10^{18}(1800\,\rm K/T_{\rm sub})^{2.8}~L_{\rm d,46}^{1/2},
\label{eq:R-IR}
\end{eqnarray}
where we take $T_{\rm sub}=1200$\,K as the sublimation temperature of the graphite grains. It reprocesses a fraction $f_{\rm IR}$ into the infrared.

The energy density of the dust torus is then given by
\begin{eqnarray}
u(\epsilon_{\star},\Omega_{\star}) &=& \zeta_{\rm IR}\frac{2(m_{\rm e}c^{2})}{\lambda_{\rm c}^{3}}\frac{\epsilon_{\star}^{3}}{\exp(\epsilon_{\star}/\Theta_{\rm IR})-1}.
\end{eqnarray}
where $\Theta_{\rm IR}=k_{\rm B}T_{\rm IR}/m_{\rm e}c^{2}$. $\zeta_{\rm IR}$ is the normalization
\begin{eqnarray}
\zeta_{\rm IR} &=& \frac{f_{\rm IR}L_{\rm d}}{4\pi R_{\rm IR}^{2}c\alpha_{\rm r}T_{\rm IR}^{4}}.
\end{eqnarray}

Finally we use the formula given by \cite{2009ApJ...692...32D} to calculate the observed EC flux.

\subsection{Absorption of BLR photons to $\gamma$-rays}
The $\gamma$-rays are absorbed via photon-pair production process $\gamma+\gamma\to e^{+}+e^{-}$.
Before calculating the absorption to $\gamma$-rays, we first present the photon number density associated with the emission lines and the diffuse continuum
\begin{eqnarray}
n_{\rm line}(\epsilon_{\star},\theta_{\star};R) &=& \frac{N_{\rm \epsilon_{\star}}}{N_{\rm \epsilon_{\star},tot}}\frac{I_{\rm line}(R,\theta_{\star})}{\epsilon_{\star}~m_{\rm e}c^{3}},
\nonumber\\
n_{\rm cont}(\epsilon_{\star},\theta_{\star};R) &=& n(R,\theta_{\star})~n_{\rm bb}(\epsilon_{\star},T_{\rm D}(R_{\rm in})),
\end{eqnarray}
where $N_{\epsilon_{\star}}$, $N_{\rm \epsilon_{\star},tot}$ are the monochromatic and total line ratios relative to the Ly$\alpha$ ratio ($N_{\epsilon_{\star},\rm Ly\alpha}=100$), respectively.
The normalization of the diffuse continuum
\begin{eqnarray}
n(R,\theta_{\star}) &=& \frac{I_{\rm cont}(R,\theta_{\star})/c}{m_{\rm e}c^{2}\int_{\epsilon_{\star,\rm l}}^{\epsilon_{\star,\rm u}}n_{\rm bb}(\epsilon_{\star},T_{\rm D}(R_{\rm in}))\epsilon_{\star} d\epsilon_{\star}},
\nonumber\\
&\simeq& \frac{4\pi}{2.7^{4}} \frac{I_{\rm cont}(R,\theta_{\star})}{\alpha_{\rm r}c T_{\rm D}^{4}(R_{\rm in})},
\end{eqnarray}
where $n_{\rm bb}(\epsilon_{\star},T_{\rm D}(R_{\rm in}))$ is the blackbody spectrum formula
\begin{eqnarray}
n_{\rm bb}(\epsilon_{\star},T_{\rm D}(R_{\rm in})) &=& \frac{2}{\lambda_{\rm c}^{3}}\frac{\epsilon_{\star}^{2}}{\exp(\epsilon_{\star}/\Theta_{\rm c})-1},
\end{eqnarray}
where
\begin{eqnarray}
\Theta_{\rm c} &=& \frac{2.7k_{\rm B}T_{\rm D}(R_{\rm in})}{m_{\rm e}c^{2}}.
\end{eqnarray}
It is worth noting that the characteristic temperature used in the calculation of the diffuse photon number density is determined at the innermost stable orbit. Thus,
\begin{eqnarray}
T_{\rm D}(R_{\rm in}) &=& \Big[\frac{3R_{\rm s}L_{\rm d}}{16\pi \sigma_{\rm T}\eta R_{\rm in}^{3}}\Big]^{1/4}.
\end{eqnarray}

Finally, the $\gamma$-ray optical depth is calculated by
\begin{eqnarray}
\tau_{\gamma\gamma}(\epsilon_{\star,\gamma}) &=& (2\pi)\int_{R_{\rm o}}^{R_{\rm max}}dR\int_{\epsilon_{\star,\rm l}}^{\epsilon_{\star,\rm u}}d\epsilon_{\star} \int_{\theta_{\star,\rm min}}^{\theta_{\star,\rm max}}\sin\theta_{\star} d\theta_{\star}
\nonumber\\
&\times& \sigma(\epsilon_{\star,\gamma},\epsilon_{\star},\theta_{\star})n(\epsilon_{\star},\theta_{\star};R)(1-\cos\theta_{\star}),
\end{eqnarray}
where $\epsilon_{\star}$ is the dimensionless energy of the diffuse photons of the BLR, and $n(\epsilon_{\star},\theta_{\star};R)=n_{\rm line}(\epsilon_{\star},\theta_{\star};R)+n_{\rm cont}(\epsilon_{\star},\theta_{\star};R)$. In the calculation, the diffuse photon frequencies are comprised between $10^{12}$\,Hz to $10^{17}$\,Hz. $R_{\rm o}$ is the position of the $\gamma$-ray emission region at a given moment, $R_{\rm max}$ is assumed to be the maximum distance above which the absorption effect can be neglected, which depends on the central BH mass and luminosity of the accretion disc. In the paper, for the specified BH mass and luminosity of FSRQ 3C 454.3, we take $R_{\rm max}=5$\,pc.

\section{Application to 3C 454.3}
To reduce free parameters for modeling the SEDs, we give some parameters as follows:

(1) The central BH mass of 3C 454.3 is $M_{\rm BH}=5\times 10^{8}$\,$M_{\odot}$ \citep{2011MNRAS.410..368B}, and the semi-aperture angle $\theta_{\rm v}$ of the jet is $1.2^{\circ}$ \citep{2005AJ....130.1418J}.

(2) The size of the emitting region $R_{\rm b}$ is determined by the observed variability timescale during the outbursts. The jet is assumed to expand in a conical shape, $R_{\rm b}$ being related to the distance $r_{\rm b}$ from the BH by $R_{\rm b}=r_{\rm b}\tan(\theta_{\rm v})$.

(3) The luminosity of the accretion disc is fixed at $5\times 10^{46}$\,erg~s$^{-1}$ for all the outbursts.

(4) The accretion disc extends from $R_{\rm in}=3$\,$R_{\rm s}$ to $R_{\rm out}=500$\,$R_{\rm s}$. Whereas the inner and outer radius of the BLR are
$r_{\rm BLR,i}=2.5\times 10^{3}$\,$R_{\rm s}$, $r_{\rm BLR,o}=5\times 10^{4}$\,$R_{\rm s}$, respectively.

(5) The spectral index $s_{1}$ of the electron energy distributions is fixed at 2.1, this value is often claimed in the literature for the particles undergoing first-order Fermi acceleration at relativistic shocks \citep{1998PhRvL..80.3911B,2000ApJ...542..235K,2001MNRAS.328..393A}.
The maximum electron energy is fixed at $\gamma_{\rm max}^{\prime}=3\times 10^{5}$ for all the outbursts.

\begin{figure}
  \centerline{
  \FigureFile(120mm,80mm){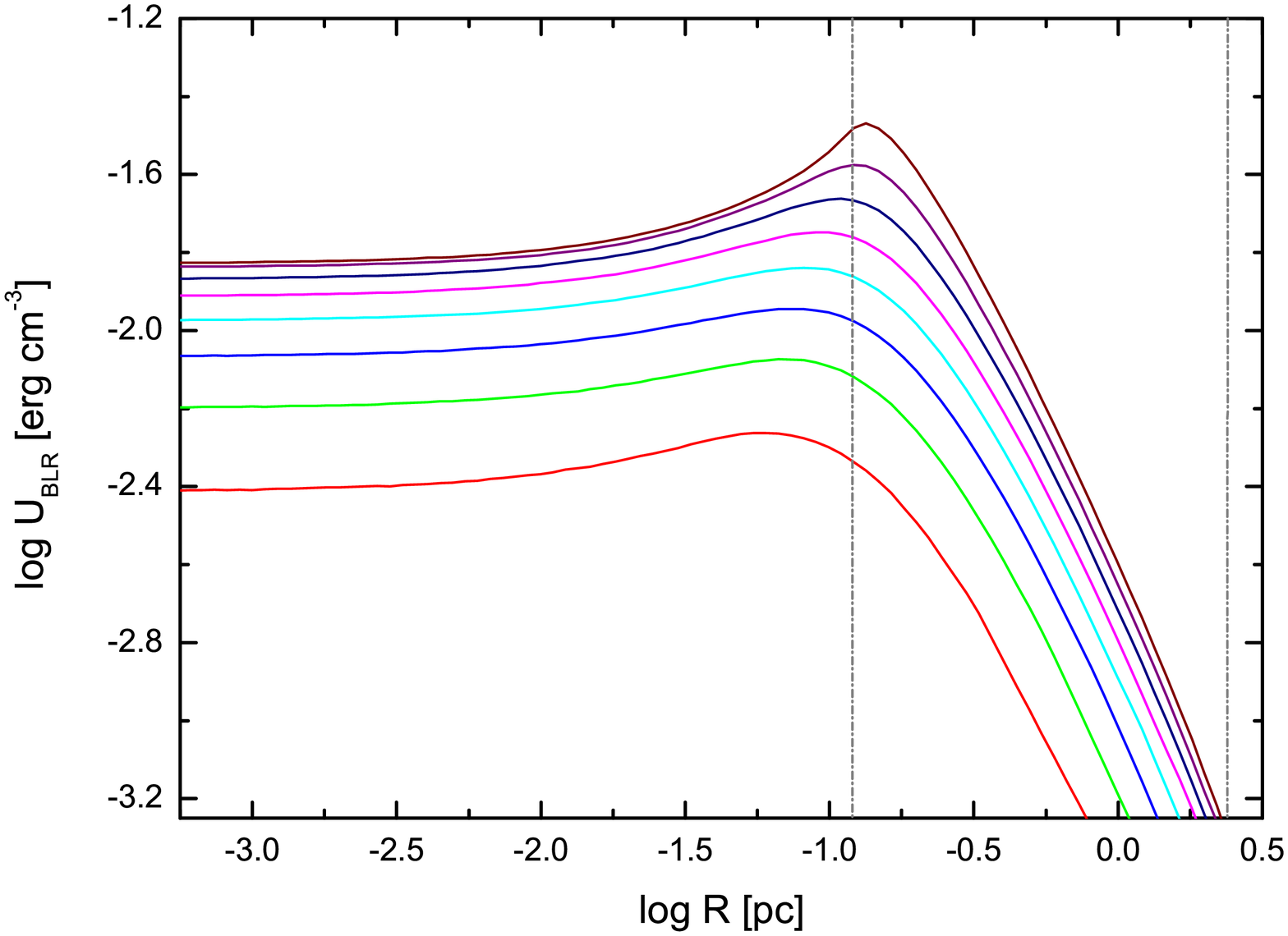}}
  \caption{Energy density $U_{\rm BLR}$ produced by a ``flat" BLR as a function of radial distance $R$ along the jet axis. The $f_{\rm line}$ and the $\tau_{\rm BLR}$ are assumed equal to 0.1. From bottom to top, the corresponding aperture angles $\alpha$ are $15^{\circ}$, $25^{\circ}$, $35^{\circ}$, $45^{\circ}$, $55^{\circ}$, $65^{\circ}$, $75^{\circ}$, $85^{\circ}$, respectively. The two vertical lines are inner and outer radius (from left to right). }
\label{fig:Colored-U_R}
\end{figure}

We calculate the contributions from the ``flat" BLR to the diffuse energy density versus the radial distance $R$ from BH, shown in Fig.\ref{fig:Colored-U_R}, in which we assume that the covering factors $f_{\rm line}$ and $\tau_{\rm BLR}$ of the BLR for emission lines and continuum (see Lei14) are equal to 0.1, i.e., the total covering factor is given by $f_{\rm BLR}=f_{\rm line}+\tau_{\rm BLR}=0.2$. Different line corresponds to different $\alpha$, from bottom to top, they are $15^{\circ}$, $25^{\circ}$, $35^{\circ}$, $45^{\circ}$, $55^{\circ}$, $65^{\circ}$, $75^{\circ}$, $85^{\circ}$, respectively. As can be seen, the energy density is almost constant inside the cavity of the BLR except for the region near the $r_{\rm BLR,i}$. The energy density then declines beyond the $r_{\rm BLR,i}$.

\begin{figure}
  \centerline{
  \FigureFile(120mm,80mm){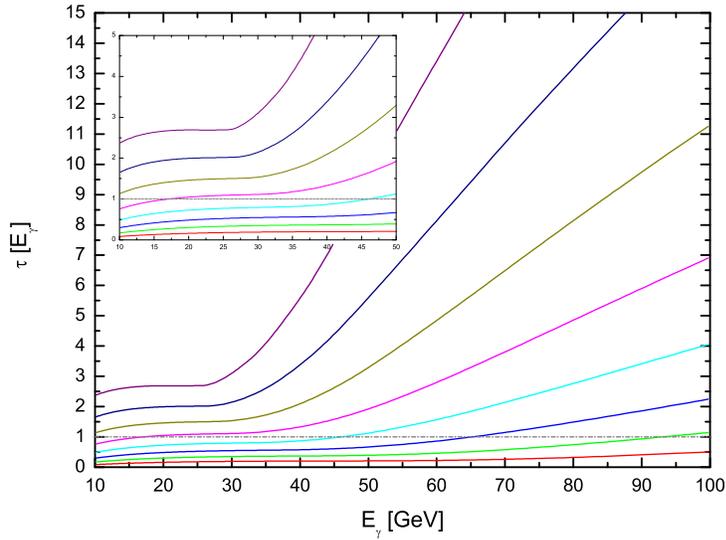}}
  \caption{Optical depth $\tau$ as a function of $\gamma$-ray photon energies $E_{\gamma}$ for different aperture angle $\alpha$, where the location of the $\gamma$-ray emission region is located at 0.2\,pc. The $f_{\rm line}$ and the $\tau_{\rm BLR}$ are assumed equal to 0.1. From bottom to top, the corresponding aperture angles $\alpha$ are $15^{\circ}$, $25^{\circ}$, $35^{\circ}$, $45^{\circ}$, $55^{\circ}$, $65^{\circ}$, $75^{\circ}$, $85^{\circ}$, respectively. Horizontal dashed line corresponds to $\tau=1$.}
\label{fig:Colored-tau_E0.2}
\end{figure}

Table \ref{tab:tab-1} presents the size of the emitting region $R_{\rm b}$, according to the assumption that the distance $r_{\rm b}$ from the BH relates to the $R_{\rm b}$ by $r_{\rm b}=R_{\rm b}/\tan(\theta_{\rm v})$, this could give a upper limit to $r_{\rm b}$. As such, most emission regions would be at position $\gtrsim 0.2$\,pc, i.e., $r_{\rm BLR,i}< r_{\rm b} <r_{\rm BLR,o}$. Fig.\ref{fig:Colored-tau_E0.2} demonstrates the absorption of the diffuse photon field from a ``flat" BLR to the $\gamma$-rays with energies from 10\,GeV up to 100\,GeV assuming the emitting region at position $R_{\rm o}=0.2$\,pc.
We find that the $\gamma$-ray photons above tens of GeV will be severely absorbed by the BLR with a spherical structure ($\alpha\simeq 85^{\circ}$ stands roughly for spherical structure), implying that at this position a ``flat" structure is required. Specifically, the $\gamma$-ray photon with energy of 50\,GeV is not absorbed for $\alpha \lesssim 45^{\circ}$. For $\alpha \lesssim 45^{\circ}$, the inset shows a spectrum from 10\,GeV to 50\,GeV, which is concave at the range of $25-30$\,GeV, corresponding to ``jumps" in the optical depth by the absorption of strong emission lines peaking at Ly$\alpha$ line. Therefore, in order to explain the gamma-ray spectra of 3C 454.3, we need a ``flat" BLR with $\alpha\lesssim 45^{\circ}$. Moreover, it should be emphasized that when the emitting region is close to or even beyond the $r_{\rm BLR,o}$, the assumption that the BLR has
a spherical structure remains a good approximation.
Recently, LAT observed short flares on timescales of 6 (or even 3) hours \citep{2010MNRAS.405L..94T,2010ApJ...721.1383A,2011ApJ...733L..26A,2012ApJ...758...72W}. This puts robust constraints to the
size (and consequently the location) of the $\gamma$-ray emitting region. As shown in Fig.\ref{fig:Colored-tau_E0.2}, this poses strong challenges to our scenario, and we will propose a scenario that hopes to solve this challenge in the following.

One of the dramatic characteristics of the blazars is that their SEDs show dual-hump structure in  $\nu-\nu F_{\nu}$ representation, e.g., a low energy component and a high energy component. In comparison to typical high-frequency-peaked BL Lac objects, the SED of FSRQs is more complicated. The high energy hump could be further divided into two components, with one peak at X-rays and the other in the GeV band. The former may originate from SSC or Compton scattering of disc radiations, and the latter from Compton scattering of BLR or dust torus radiations, or the sum of the both. Moreover, due to the synchrotron self-absorption in radio bands, the lack of the soft $\gamma$-ray data, as well as the low sensitivity in GeV $\gamma$-rays, the three peaks cannot be well constrained by the observations. For modeling our collected 26 quasi-simultaneous SEDs obtained by the multi-wavelength campaigns spanned from 2007 July to 2011 January, we fix the aperture angle $\alpha$ at $45^{\circ}$ as an upper limit.
Based on the valid expressions in Thomson regime presented in Appendix, which relates the model parameters to the observables, and external radiation energy density $u_{\rm ext}$, together with spectral indexes $\alpha_{1}$ and $\alpha_{2}$, the values of the Doppler factor $\delta_{\rm D}$ and the magnetic field strength B are determined. Because the third component could be a sum of different radiation components, $\delta_{\rm D}$ and B are not accurate, and they are only a set of ``baseline parameters". Using this set of parameters, we model the collected quasi-simultaneous SEDs. We assume that the covering factor for the BLR and the dust torus take the same values during each multiwavelength campaign, whereas the covering factor could be different for different campaign. The results are shown in Fig.\ref{fig:SED-1} $-$ Fig.\ref{fig:SED-4}, the corresponding parameters are reported in Table \ref{tab:tab-1}. As such, the low-energy component is well explained by synchrotron radiation. For the high-energy component, the first hump is produced by SSC emission, the second one by both EC scattering of the dust torus and the BLR radiations. It is worth noting that there is some degeneracy in the choice of the model parameters, hence the set of parameter values are not unique. Our work only provide an attempt to explore the structure and properties of the BLR of 3C 454.3.

To explore the SED properties based on the modeling results, we calculate typical peak frequencies.
In the EC emission, the $\gamma$-ray spectral component is expected to peak at the observed energy
\begin{eqnarray}
\nu_{\rm obs}&=& 5.5\times 10^{-24} \frac{\delta_{\rm D}\Gamma_{\rm j}}{1+z}{\gamma_{\rm br}^{\prime}}^{2}\nu_{\rm ext}\,\rm GeV,
\end{eqnarray}
where $\nu_{\rm ext}$ is the characteristic frequency of the external radiation field, corresponding to  $\nu_{\rm Ly\alpha}=2.47\times 10^{15}$\,Hz and $\nu_{\rm IR}=3\times 10^{13}$\,Hz for the BLR and the dust torus, respectively. In the comoving frame, the external photon frequency is observed at $\nu_{\rm ext}^{\prime}\simeq 2\Gamma_{\rm j}\nu_{\rm ext}$. If the scattering take place in Thomson regime, we can estimate two maximum energies given by \citep{2009MNRAS.397..985G}
\begin{eqnarray}
\nu_{\rm KN}^{\rm BLR} &=& 15\frac{\delta_{\rm D}}{\Gamma_{\rm j}(1+z)}\,\rm GeV,
\nonumber\\
\nu_{\rm KN}^{\rm IR} &=& 1.2\frac{\delta_{\rm D}}{\Gamma_{\rm j}(1+z)}\,\rm TeV.
\end{eqnarray}

We also give two additional frequencies
\begin{eqnarray}
\nu_{\rm syn,13} &=& 3.7\times 10^{-7} B{\gamma_{\rm br}^{\prime}}^{2}\frac{\delta_{\rm D}}{1+z},
\nonumber\\
\nu_{\rm SSC,19} &=& 1.33\times 10^{-6}{\gamma_{\rm br}^{\prime}}^{2}\nu_{\rm syn,13},
\end{eqnarray}
where $\nu_{\rm syn,13}$, $\nu_{\rm SSC,19}$ are the synchrotron and SSC peak frequency, respectively.

Finally we estimate the powers of the Poynting flux ($P_{\rm B}$), the relativistic electrons ($P_{\rm e}$) and the cold protons ($P_{\rm p}$) during the outbursts. All powers are calculated via \citep{2008MNRAS.385..283C,2010MNRAS.402..497G}
\begin{eqnarray}
P_{\rm i}&=& \pi R_{\rm b}^{2} \Gamma_{\rm j}^{2}cU_{\rm i}^{\prime},
\end{eqnarray}
where $U_{\rm i}^{\prime}$ is the energy density of the $i$ component in the comoving frame. The electron energy density is given by $U_{\rm e}^{\prime}=m_{\rm e}c^{2}\int_{0}^{\infty} \gamma^{\prime}n_{\rm e}(\gamma^{\prime})d\gamma^{\prime}$. The power carried by protons is calculated by assuming one proton per fifty electrons \citep{2000ApJ...534..109S}.

The power taken away by radiation (including synchrotron, SSC and EC emission) is
\begin{eqnarray}
P_{\rm r}= \pi R_{\rm b}^{2} \Gamma_{\rm j}^{2}cU_{\rm r}^{\prime}=\frac{\Gamma_{\rm j}^{2}}{\delta_{\rm D}^{4}}\frac{L_{\rm tot}}{4},
\end{eqnarray}
where $U_{\rm r}^{\prime}=L_{\rm tot}^{\prime}/(4\pi R_{\rm b}^{2}c)$ is the comoving radiation energy density, $L_{\rm tot}^{\prime}$ and $L_{\rm tot}$ are the total luminosities measured in the comoving
and the observer frame, respectively. The synchrotron ($L_{\rm syn}$), SSC ($L_{\rm SSC}$) and EC ($L_{\rm EC}$) as well as the total luminosity in the observer frame are given by
\begin{eqnarray}
L_{\rm syn} &=& 4\pi d_{\rm L}^{2}(1+z)^{2}\int_{0}^{\infty} \frac{f_{\epsilon}^{\rm syn}}{\epsilon_{\rm s}^{\rm o}}d\epsilon_{\rm s}^{\rm o},
\nonumber\\
L_{\rm SSC} &=& 4\pi d_{\rm L}^{2}(1+z)^{2}\int_{0}^{\infty} \frac{f_{\epsilon_{\rm s}}^{\rm SSC}}{\epsilon_{\rm s}^{\rm o}}d\epsilon_{\rm s}^{\rm o},
\nonumber\\
L_{\rm EC} &=& 4\pi d_{\rm L}^{2}\int_{0}^{\infty}\frac{f_{\epsilon_{\rm s}}^{\rm EC}}{\epsilon_{\rm s}^{\rm o}}d\epsilon_{\rm s}^{\rm o}.
\nonumber\\
L_{\rm tot} &=& L_{\rm syn}+L_{\rm SSC}+L_{\rm EC}.
\end{eqnarray}
The estimated powers and luminosity are shown in the Table 2.

\section{Discussion}
In this paper, we collected 9 multi-wavelength campaigns including 26 quasi-simultaneous SEDs in total. From Table \ref{tab:tab-1}, we can see that: (i) The significant flaring activity is closely related to the increase of $\delta_{\rm D}$ except for two cases (Verc09 and Verc10), in which the increase of the normalization factor $n_{0}$ of the electron distribution seems to plays an important role in triggering the outburst. (ii) For the multi-wavelength campaign of Pacc10, the resulting electron minimum energy $\gamma_{\rm min}^{\prime}$ are from 200 to 240 for different SED, larger than the values required in the other multi-wavelength campaigns. Let $\gamma_{\rm min}$ takes larger value is for better fitting the steep low-energy-end SED of the SSC emission, as illustrated in Fig.\ref{fig:SED-2}. (iii) The total covering factor of the BLR ($f_{\rm BLR}=f_{\rm line}+\tau_{\rm BLR}$) is inversely related to the covering factor of the dust torus $f_{\rm IR}$, e.g., lower $f_{\rm BLR}$ corresponds to larger $f_{\rm IR}$. When BLR has lower obscuration, more disc radiation could
illuminate the dust torus, resulting in larger infrared radiation. (iv) For a small size of the emitting region (consequently short variability timescale), this seems to correspond to lower $f_{\rm BLR}$ and larger $f_{\rm IR}$. If this is the true, it implies that when the emitting region locates inside the cavity of the BLR,
the matter of the BLR disperses at larger space which is even larger than $45^{\circ}$, due to larger dispersion, this leads to lower obscuration, thus, $f_{\rm BLR}$ will be small (even $\leq 0.05$), and $f_{\rm IR}$ become large. As the emitting region goes into the BLR ($r_{\rm BLR,i}<r_{\rm b}<r_{\rm BLR,o}$), there are interactions between the emitting region and BLR clouds, and the BLR clouds will occupy smaller volume to increase $f_{\rm BLR}$. Actually, the aperture angle $\alpha=45^{\circ}$ could be an averaged angle which measures the distribution of the matter of the BLR.

\begin{figure}
  \centerline{
  \FigureFile(120mm,80mm){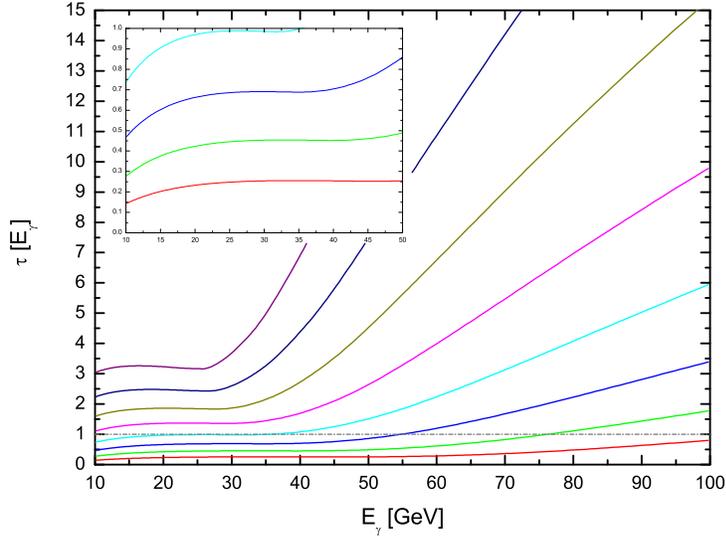}}
  \caption{Optical depth $\tau$ as a function of $\gamma$-ray photon energies $E_{\gamma}$ for different aperture angle $\alpha$, where the location of the emission region is located at $6.7\times 10^{-2}$\,pc, corresponding to 3\,h variability timescales, the $\gamma$-ray emission region is assumed to has a Lorentz factor $\delta_{\rm D}=25$. The $f_{\rm line}$ and the $\tau_{\rm BLR}$ are assumed equal to 0.05. From bottom to top, the corresponding aperture angles $\alpha$ are $15^{\circ}$, $25^{\circ}$, $35^{\circ}$, $45^{\circ}$, $55^{\circ}$, $65^{\circ}$, $75^{\circ}$, $85^{\circ}$, respectively. Horizontal dashed line corresponds to $\tau=1$.}
\label{fig:Colored-tau_E7.6-2}
\end{figure}

\begin{figure*}
\includegraphics[scale=0.28,angle=0]{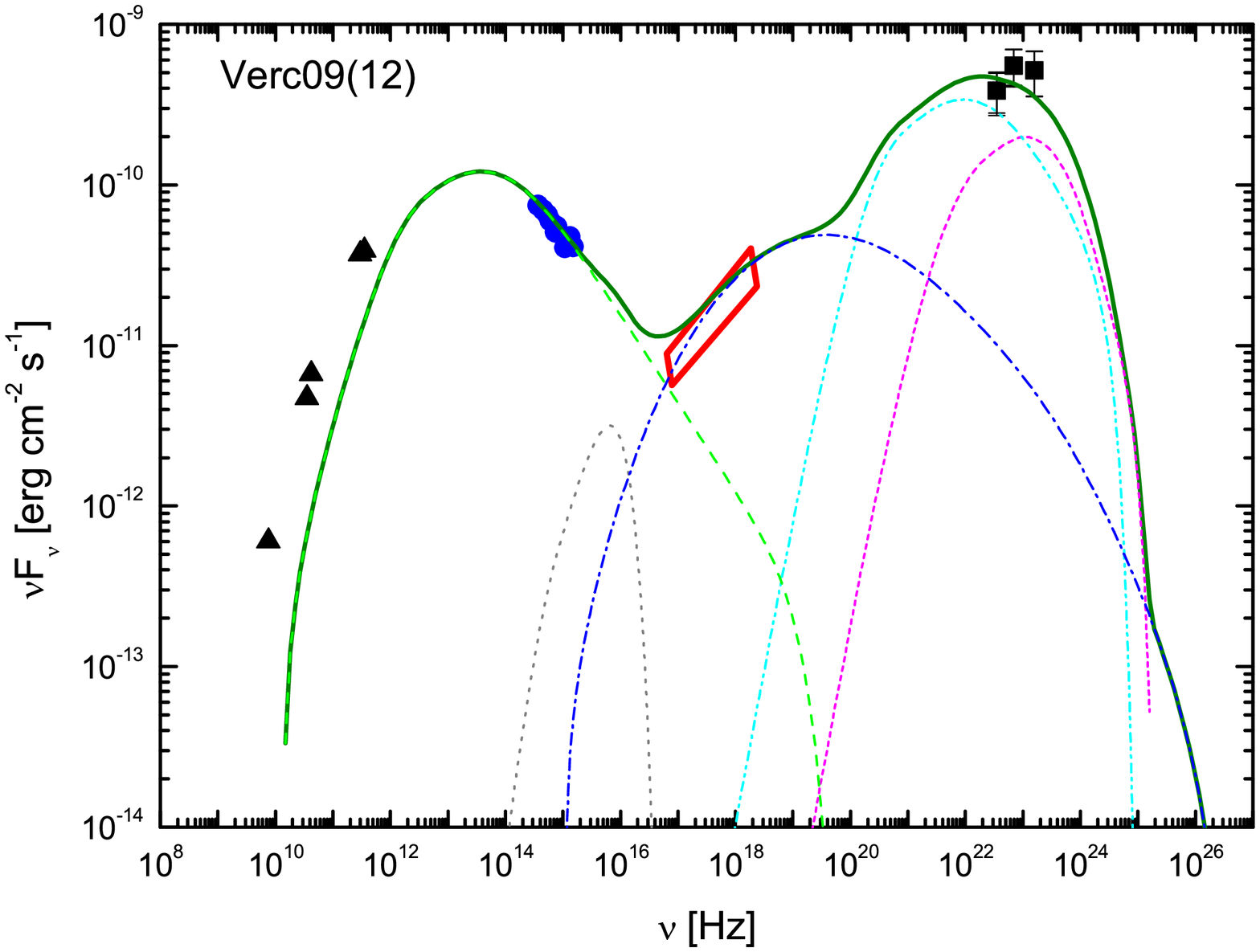}
\includegraphics[scale=0.28,angle=0]{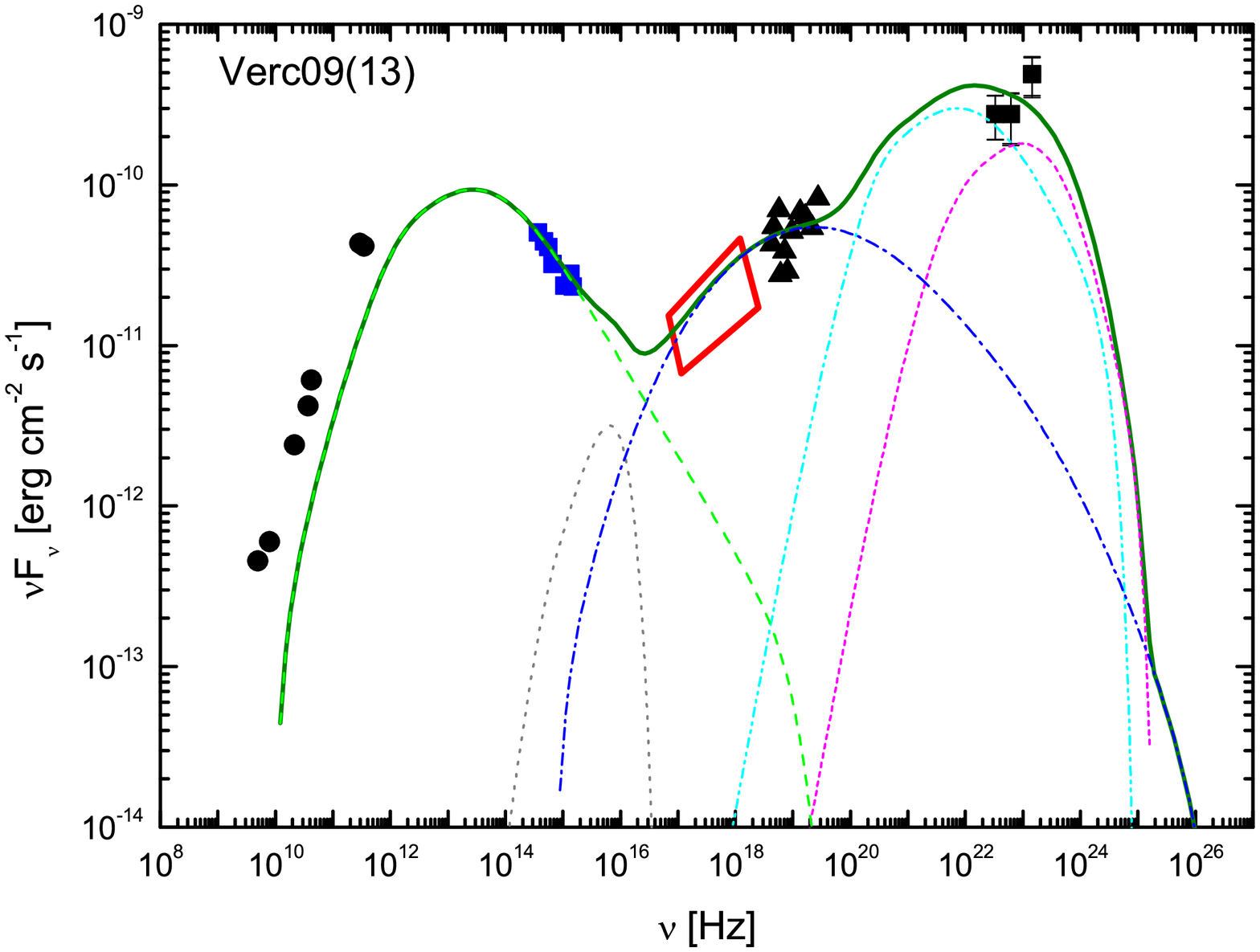}\\
\includegraphics[scale=0.28,angle=0]{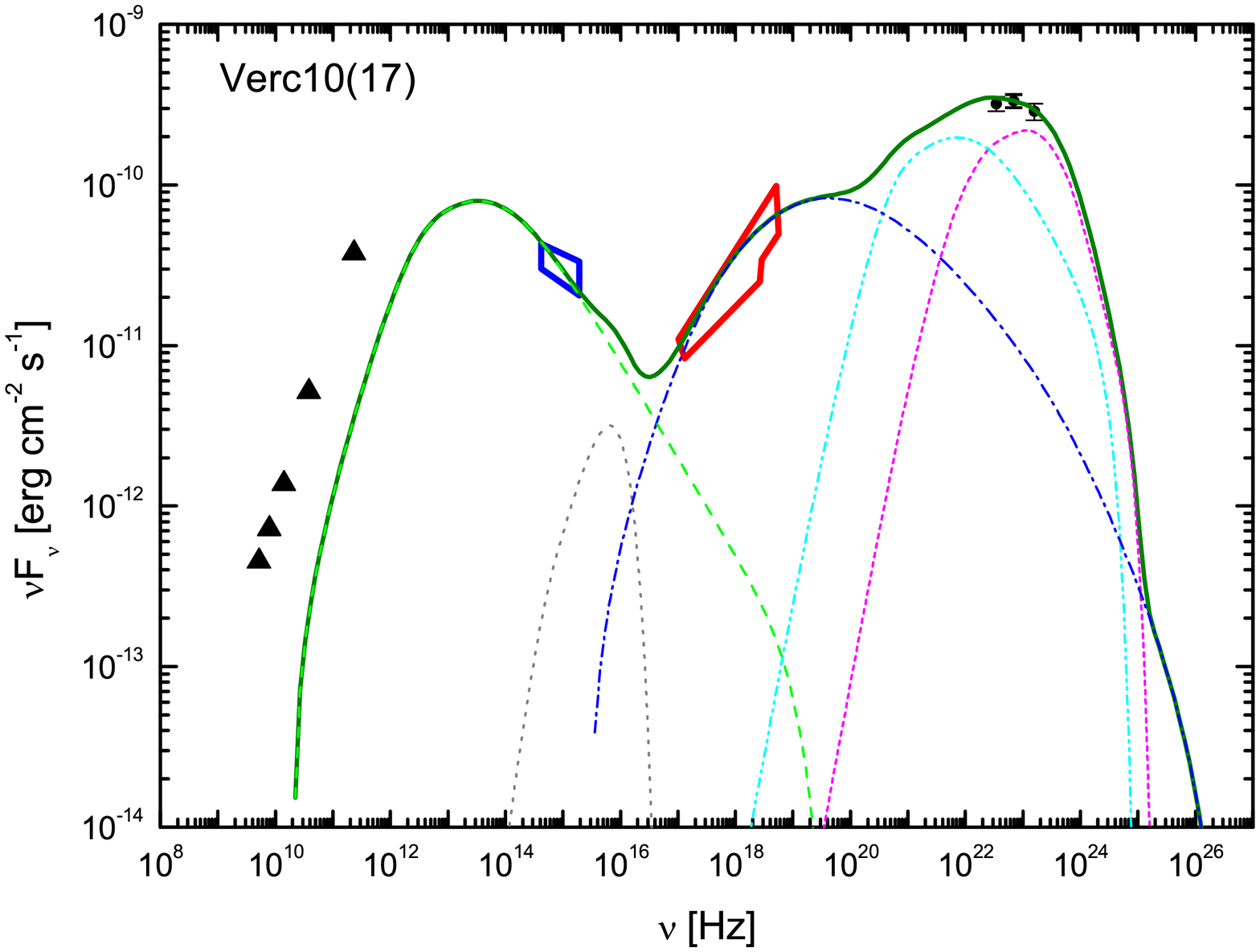}
\includegraphics[scale=0.28,angle=0]{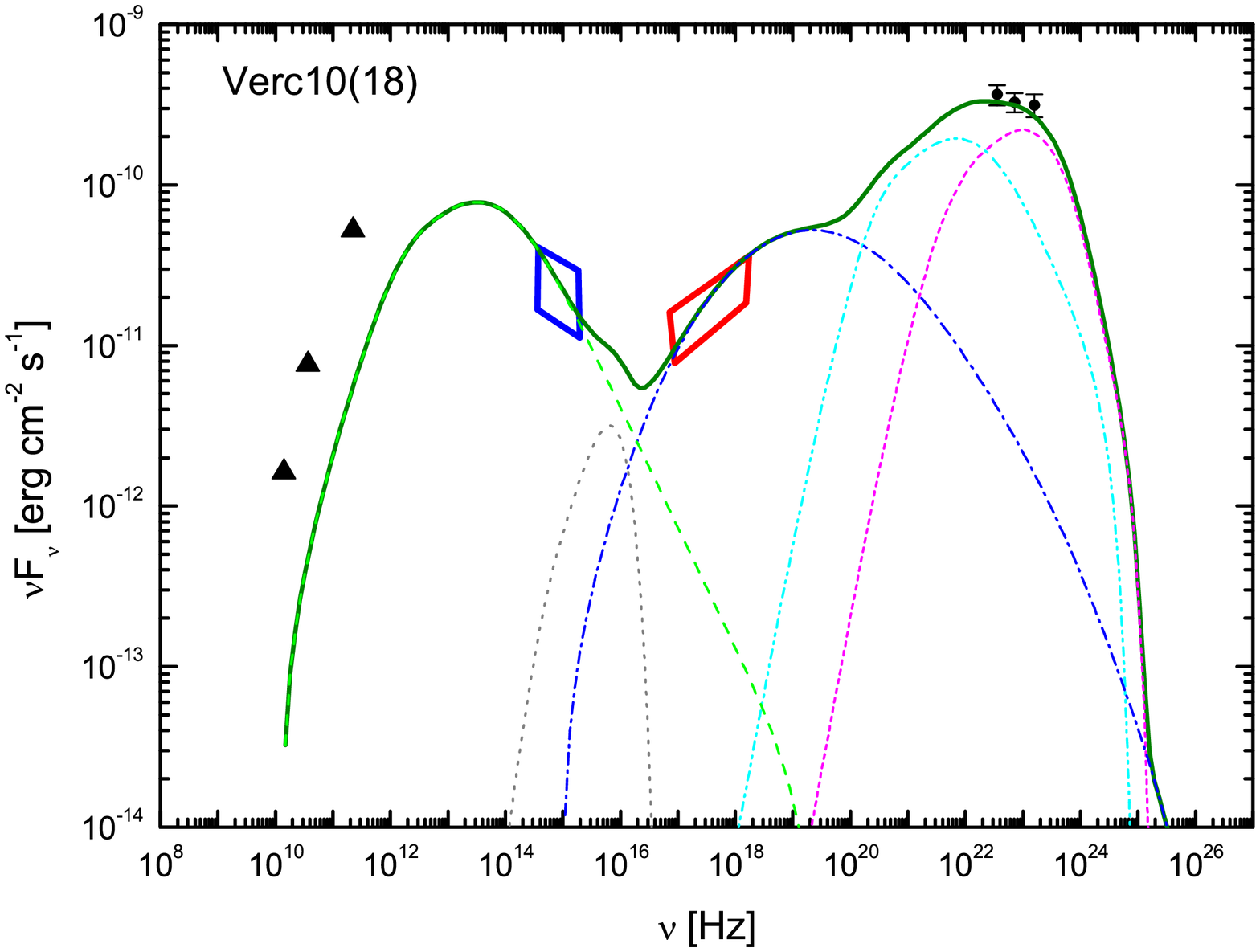}\\
\includegraphics[scale=0.28,angle=0]{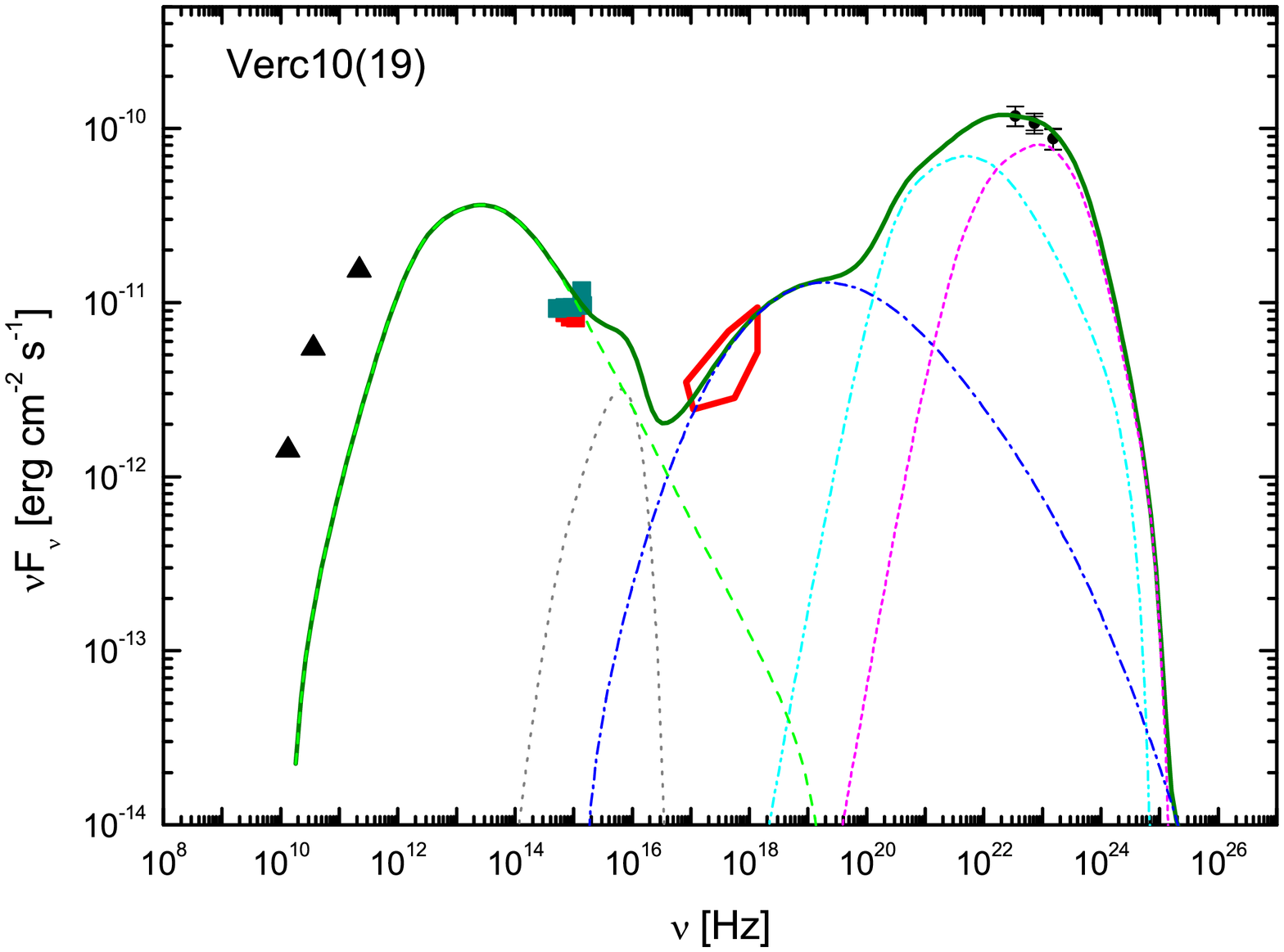}
\includegraphics[scale=0.28,angle=0]{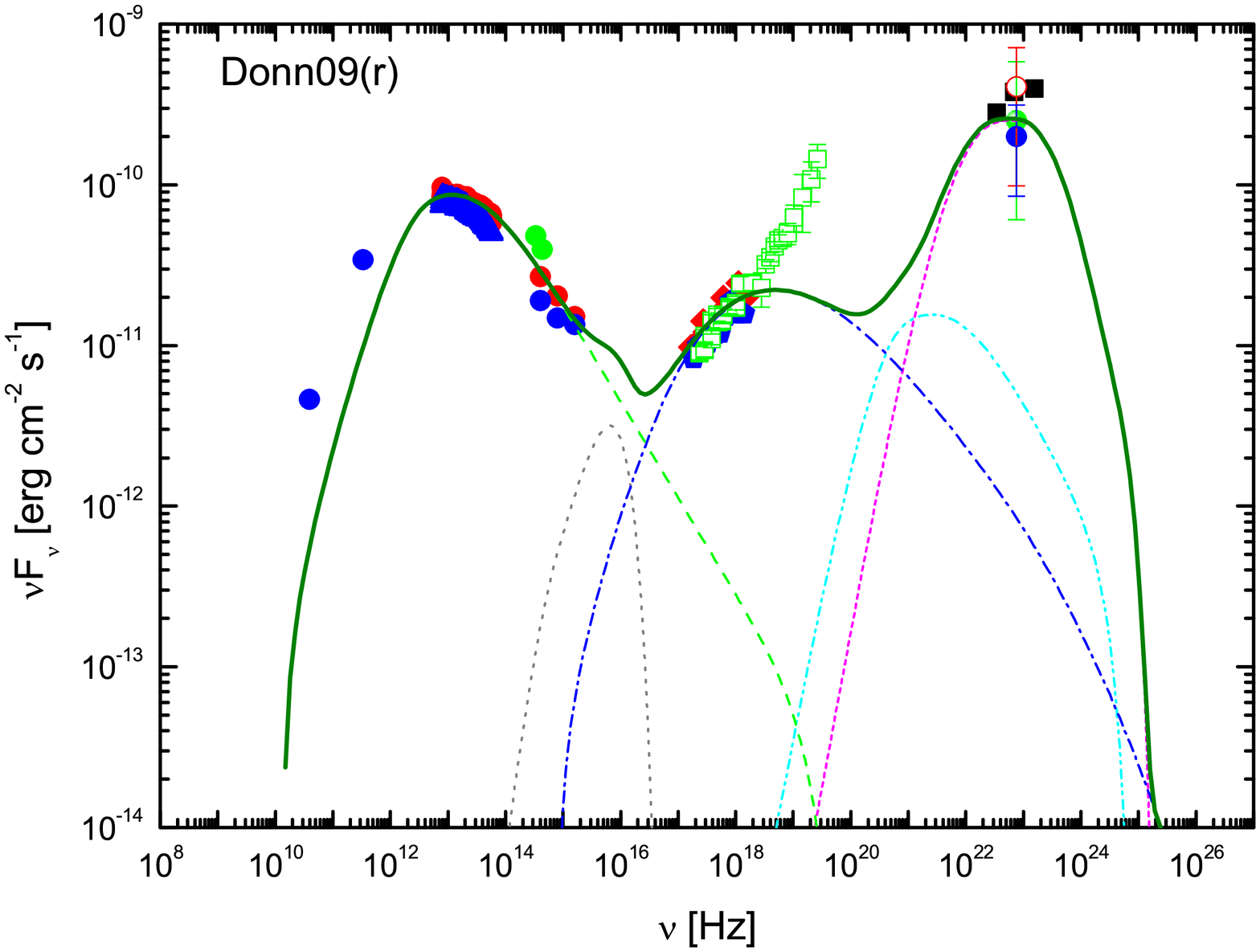}\\
\includegraphics[scale=0.28,angle=0]{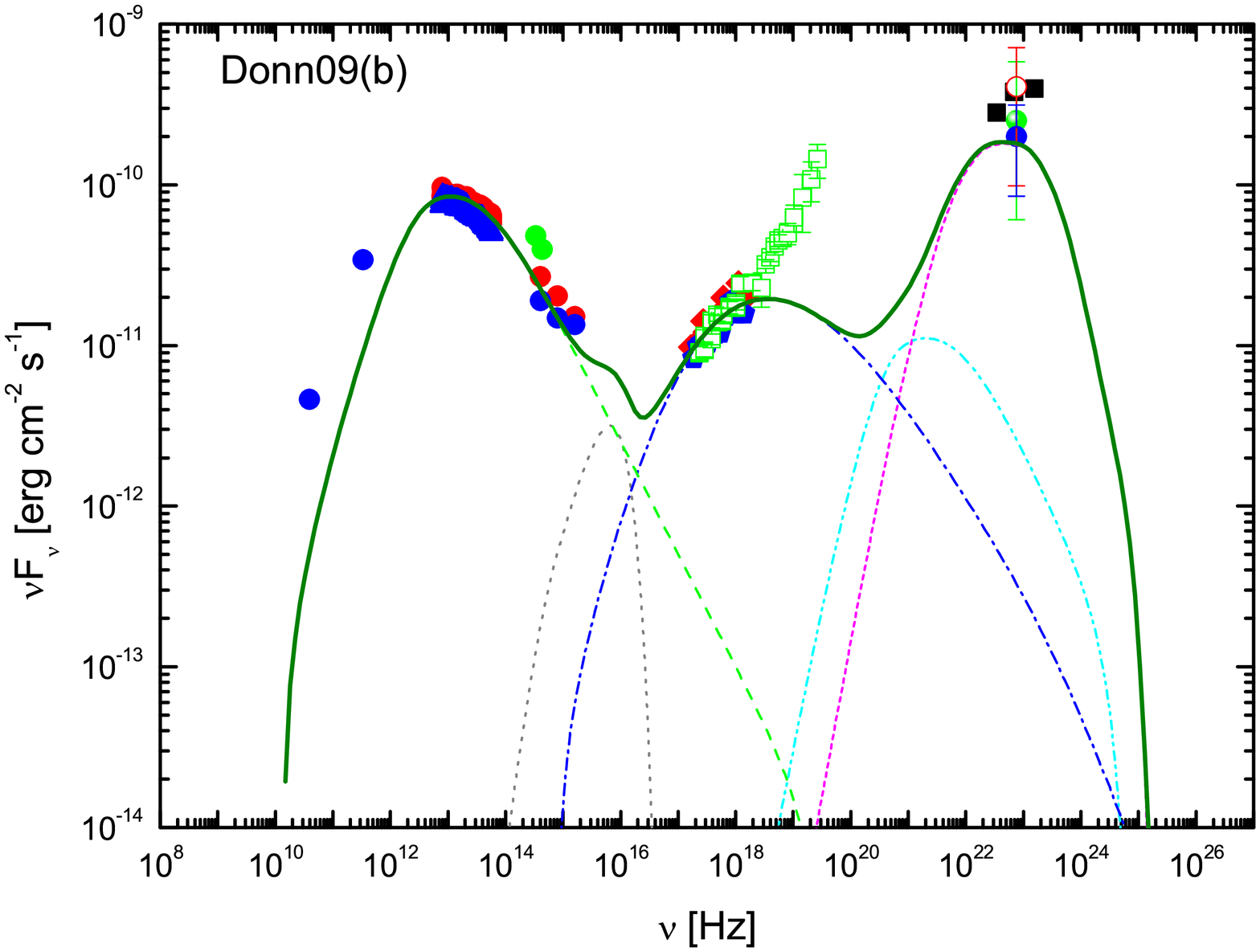}
\includegraphics[scale=0.28,angle=0]{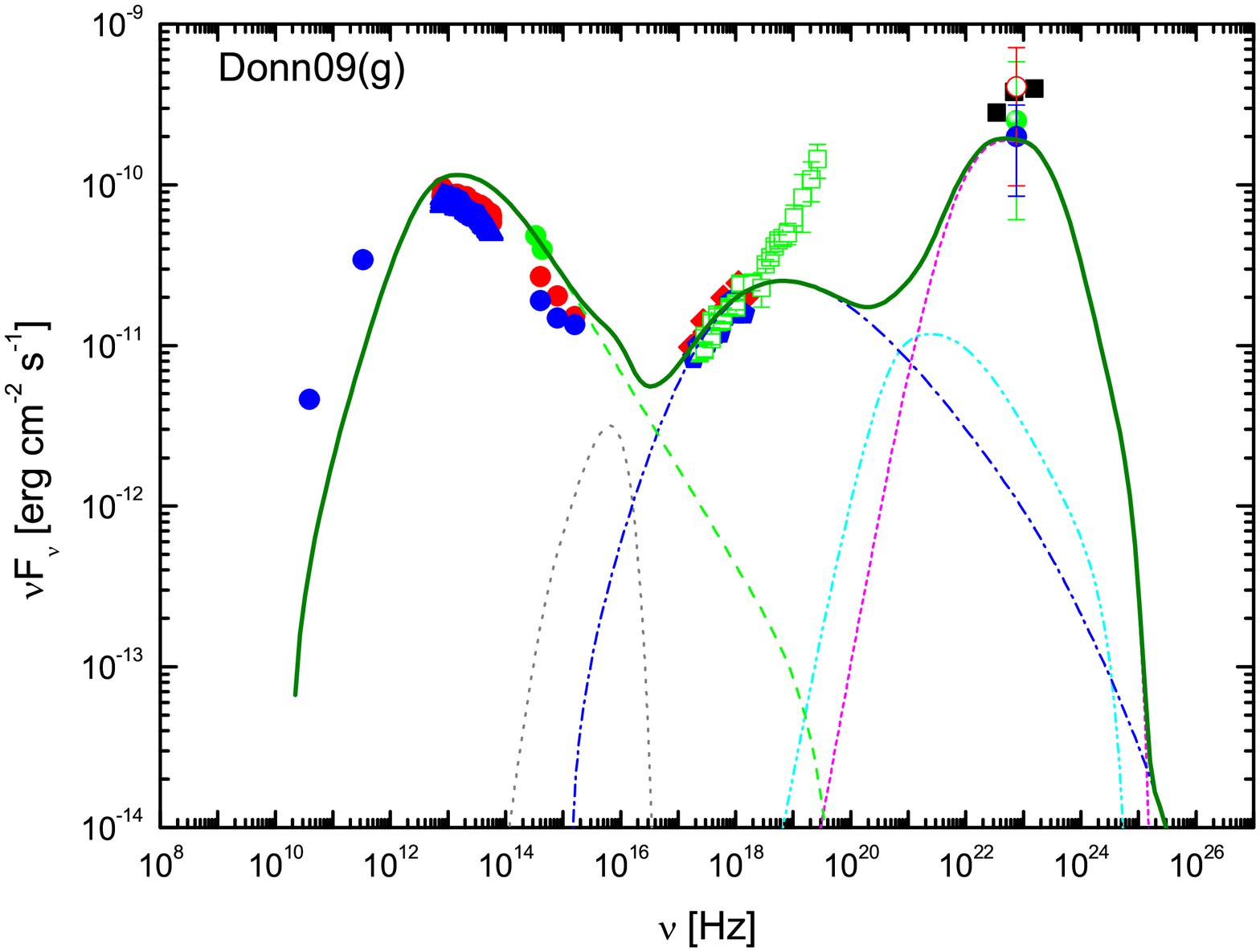}\\
\caption{Collected SEDs together with the fitting model, with the parameters listed in Table \ref{tab:tab-1}. The discrete points are the observed data, in which part X-ray data are represented with irregular polygon. Separate spectral components (thin short dashed curves) are, from left to right, the synchrotron, the accretion disc, SSC, Compton scattering of the dust torus and the BLR radiations, respectively. The thick solid line represents the superposition of all the components.}
\label{fig:SED-1}
\end{figure*}

\begin{figure*}
\includegraphics[scale=0.28,angle=0]{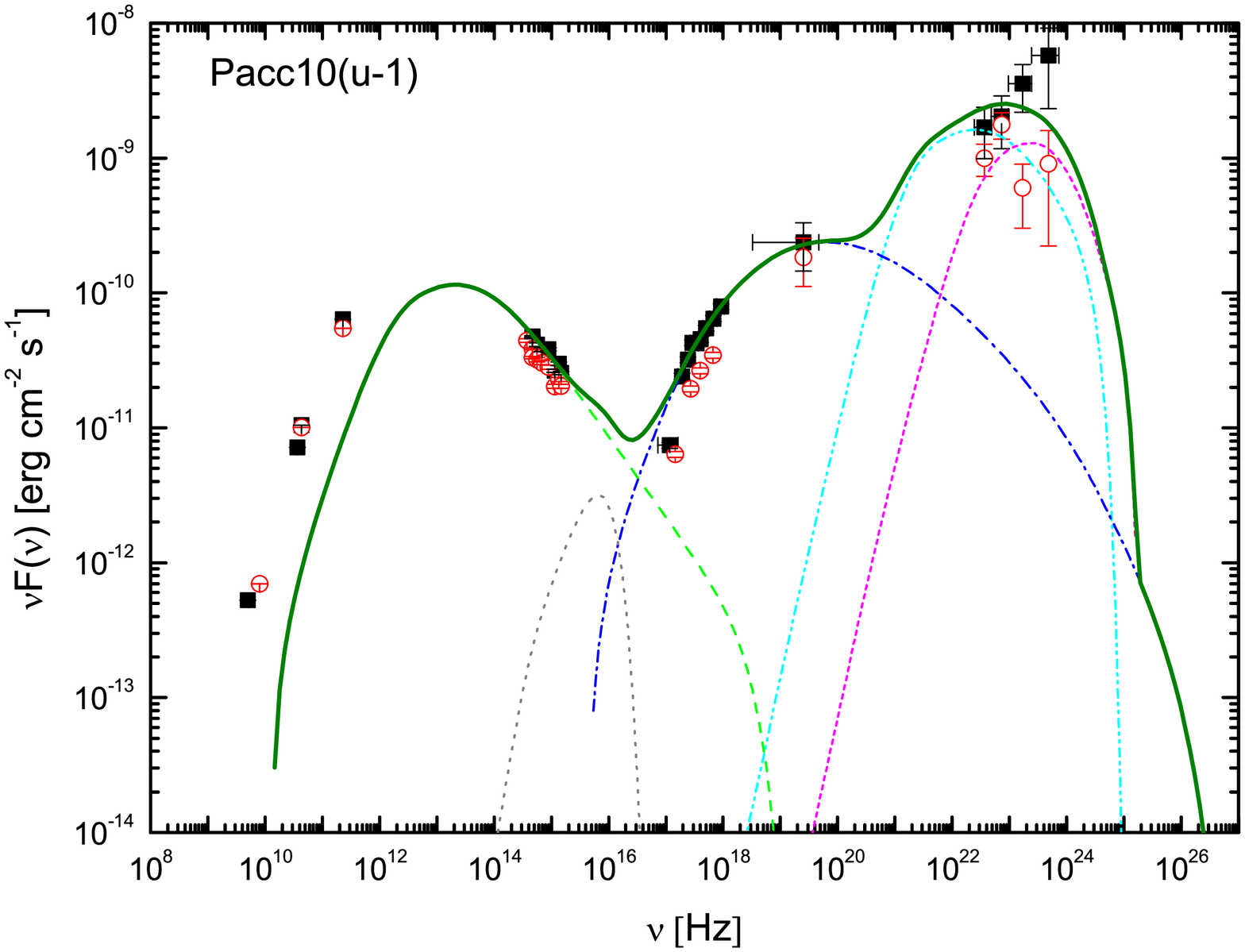}
\includegraphics[scale=0.28,angle=0]{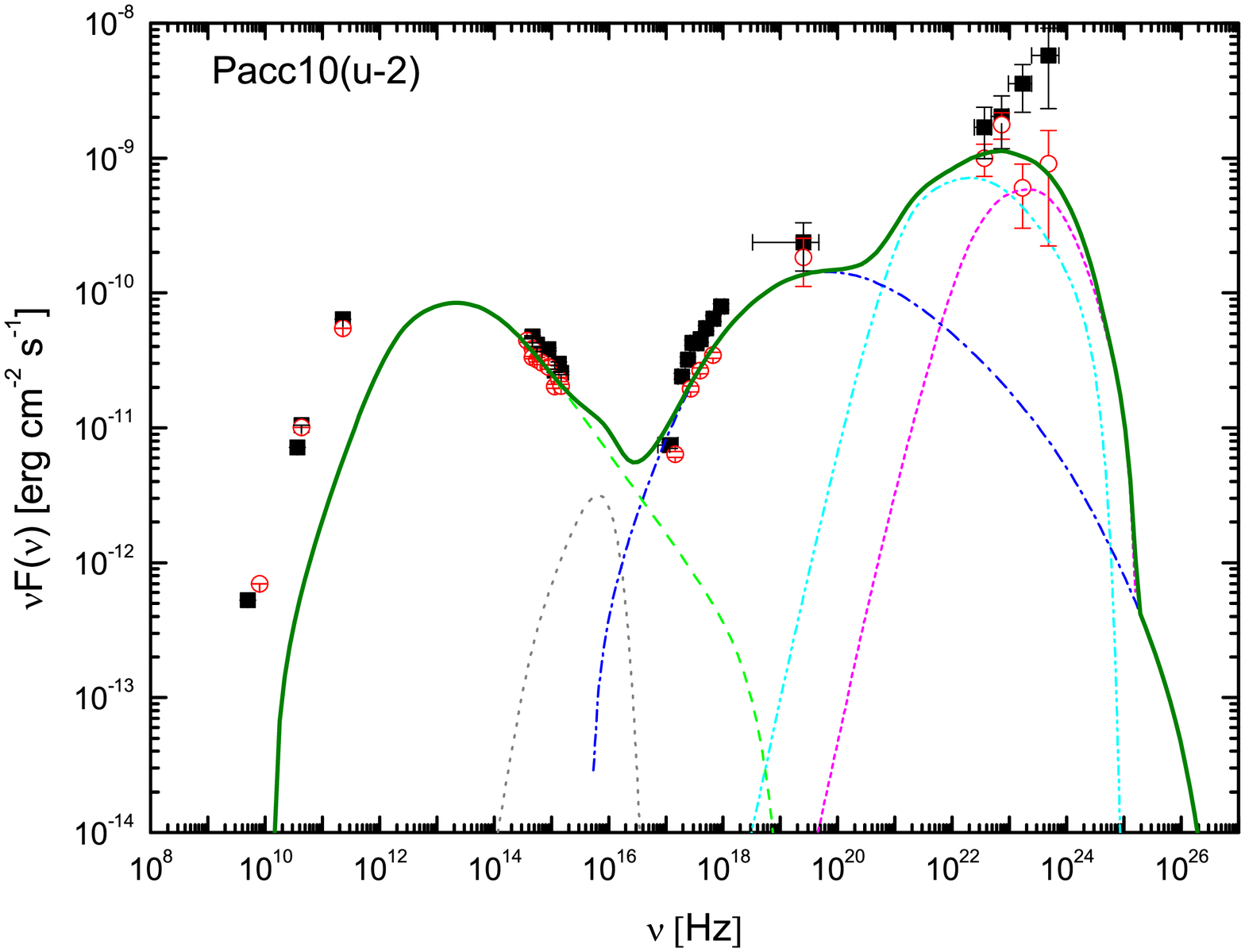}\\
\includegraphics[scale=0.28,angle=0]{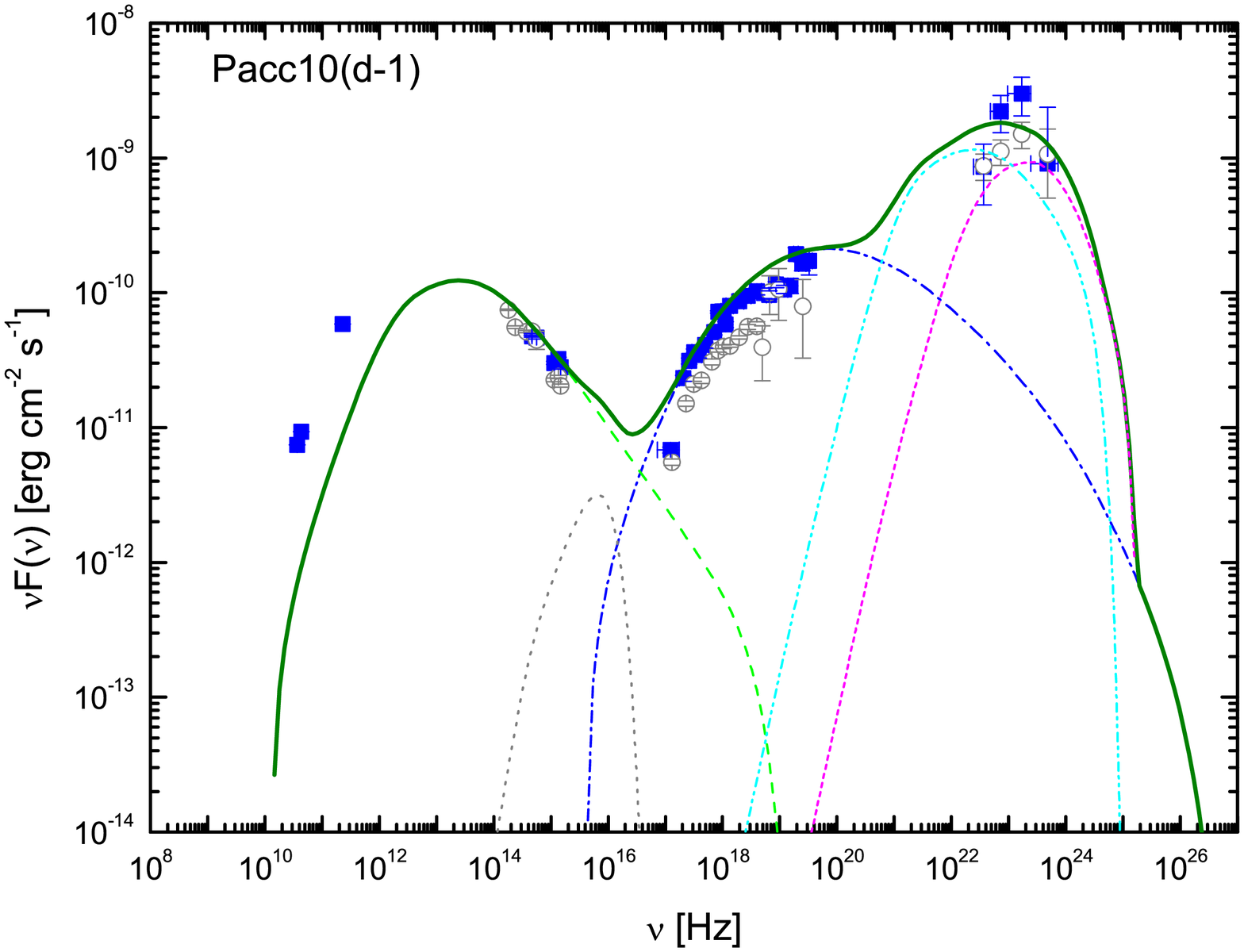}
\includegraphics[scale=0.28,angle=0]{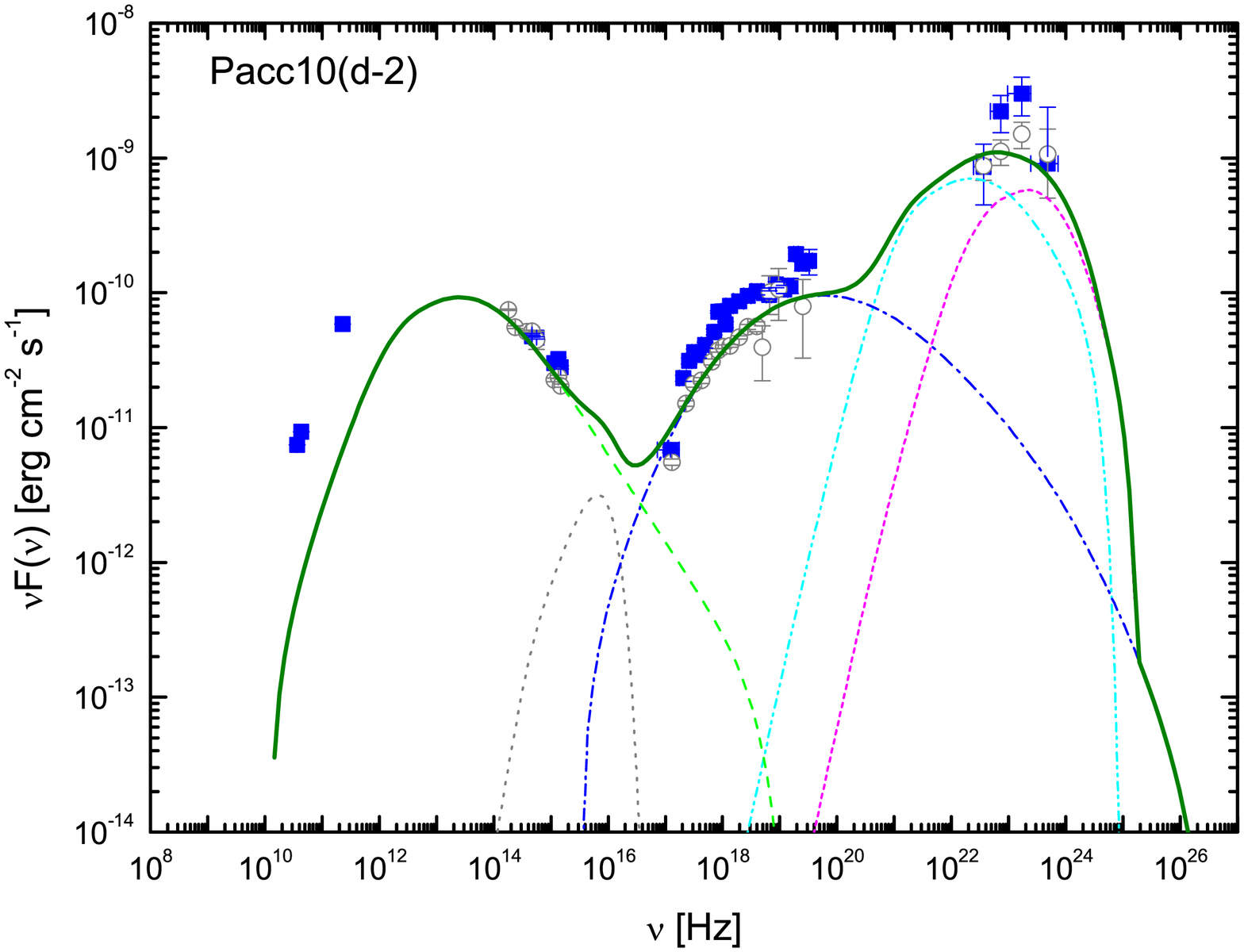}\\
\includegraphics[scale=0.28,angle=0]{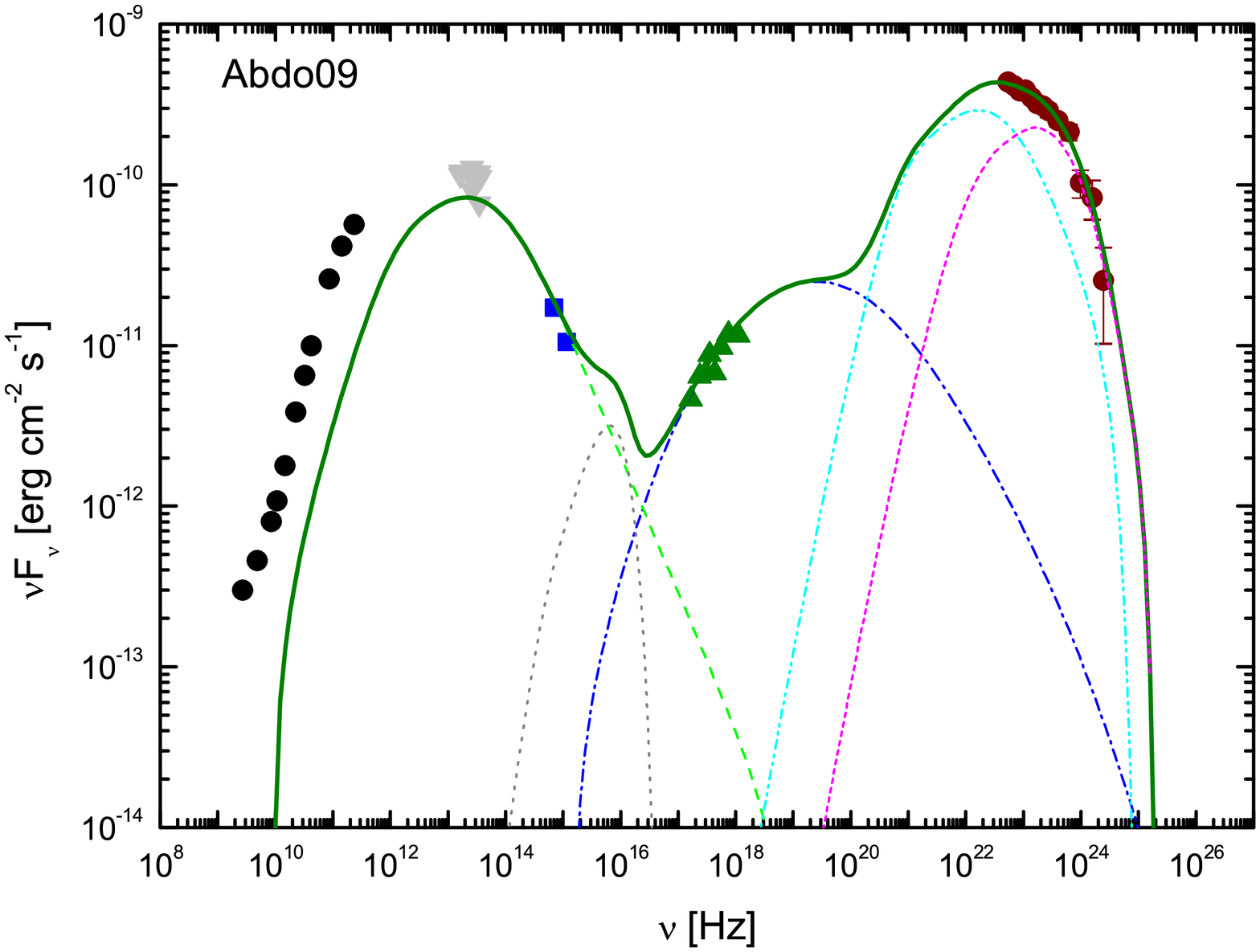}
\includegraphics[scale=0.28,angle=0]{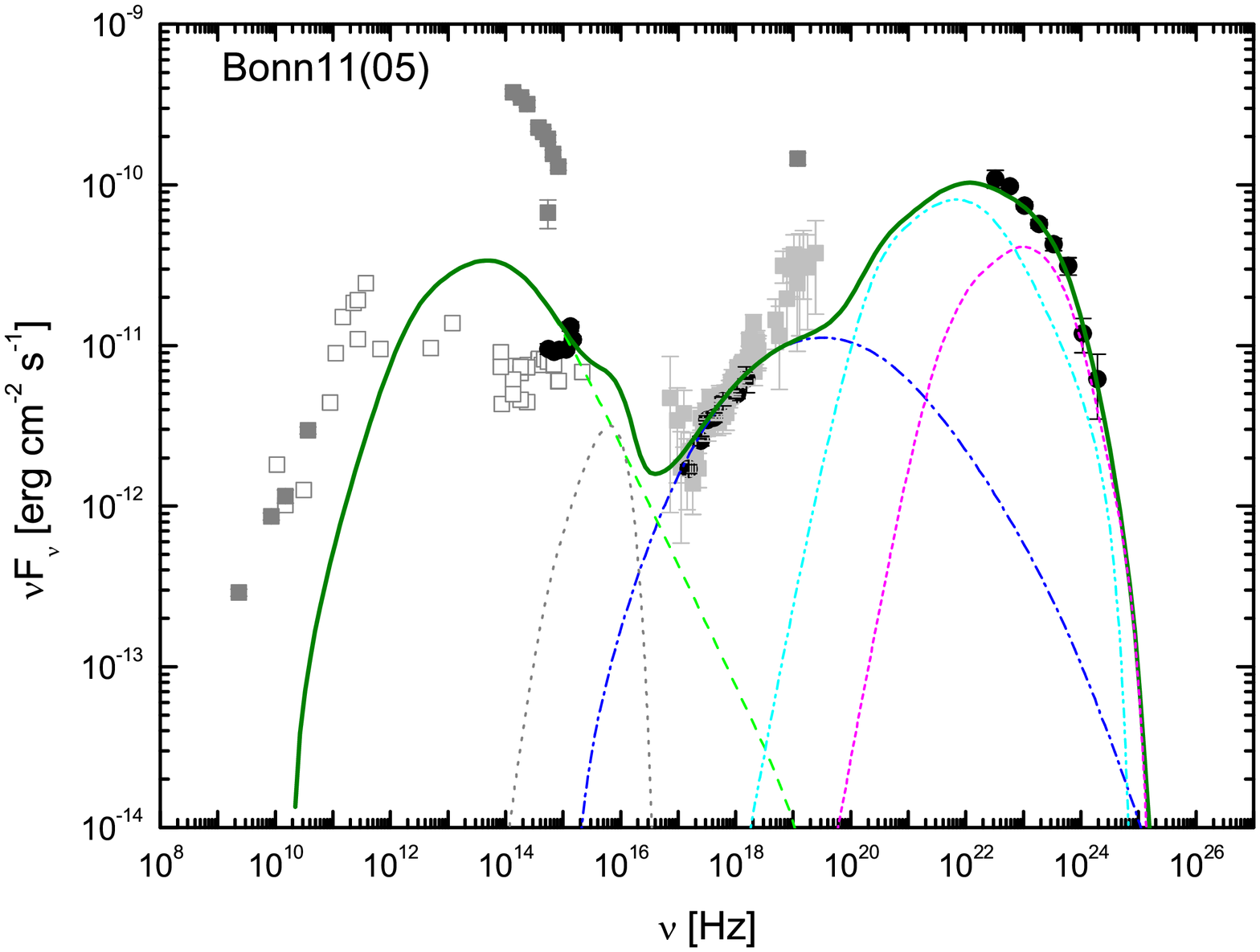}\\
\includegraphics[scale=0.28,angle=0]{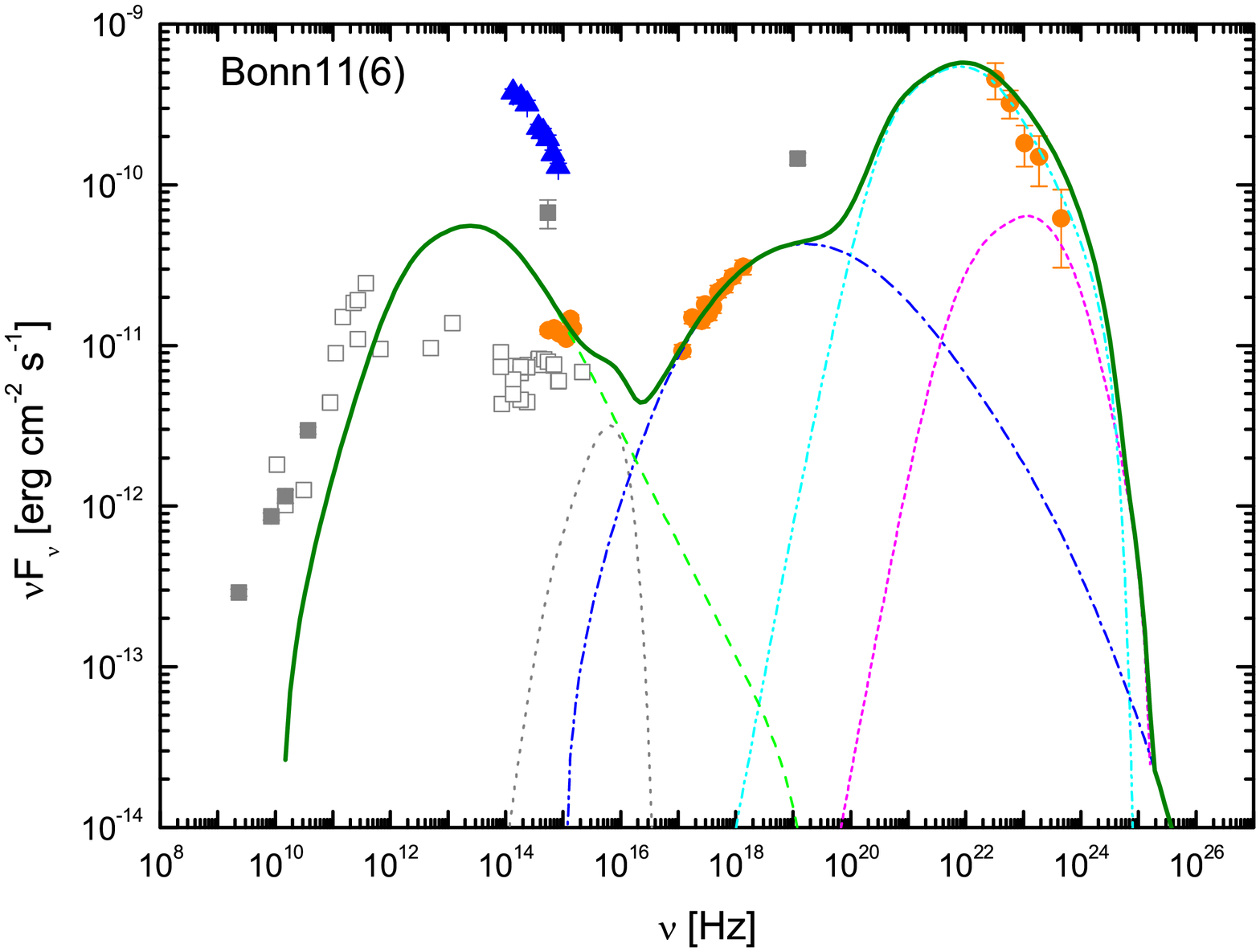}
\includegraphics[scale=0.28,angle=0]{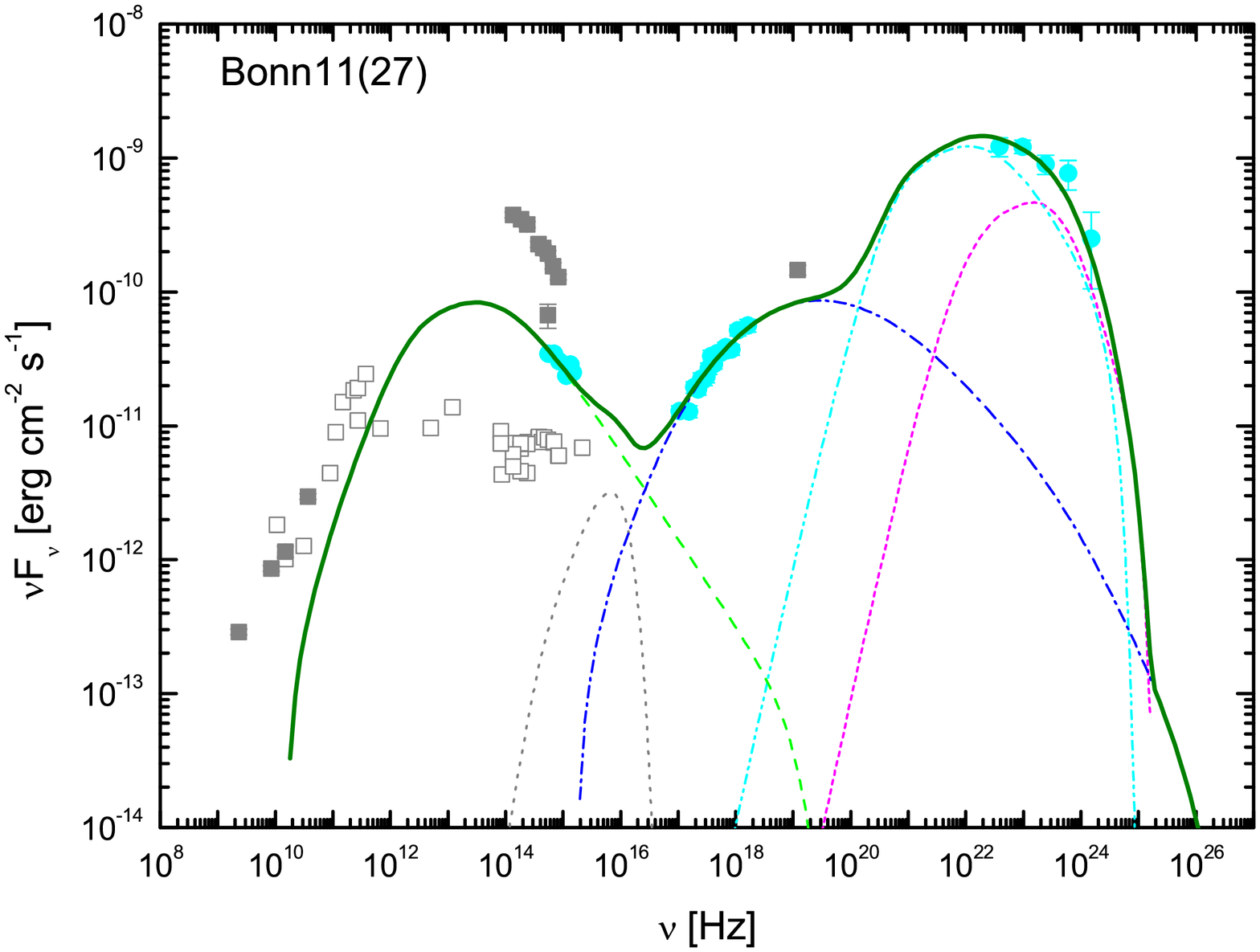}\\
\caption{Collected SEDs together with the fitting model, with the parameters listed in Table \ref{tab:tab-1}. The discrete points are the observed data. Separate spectral components (thin short dashed curves) are, from left to right, the synchrotron, the accretion disc, SSC, Compton scattering of the dust torus and the BLR radiations, respectively. The thick solid line represents the superposition of all the components.}
\label{fig:SED-2}
\end{figure*}

\begin{figure*}
\includegraphics[scale=0.28,angle=0]{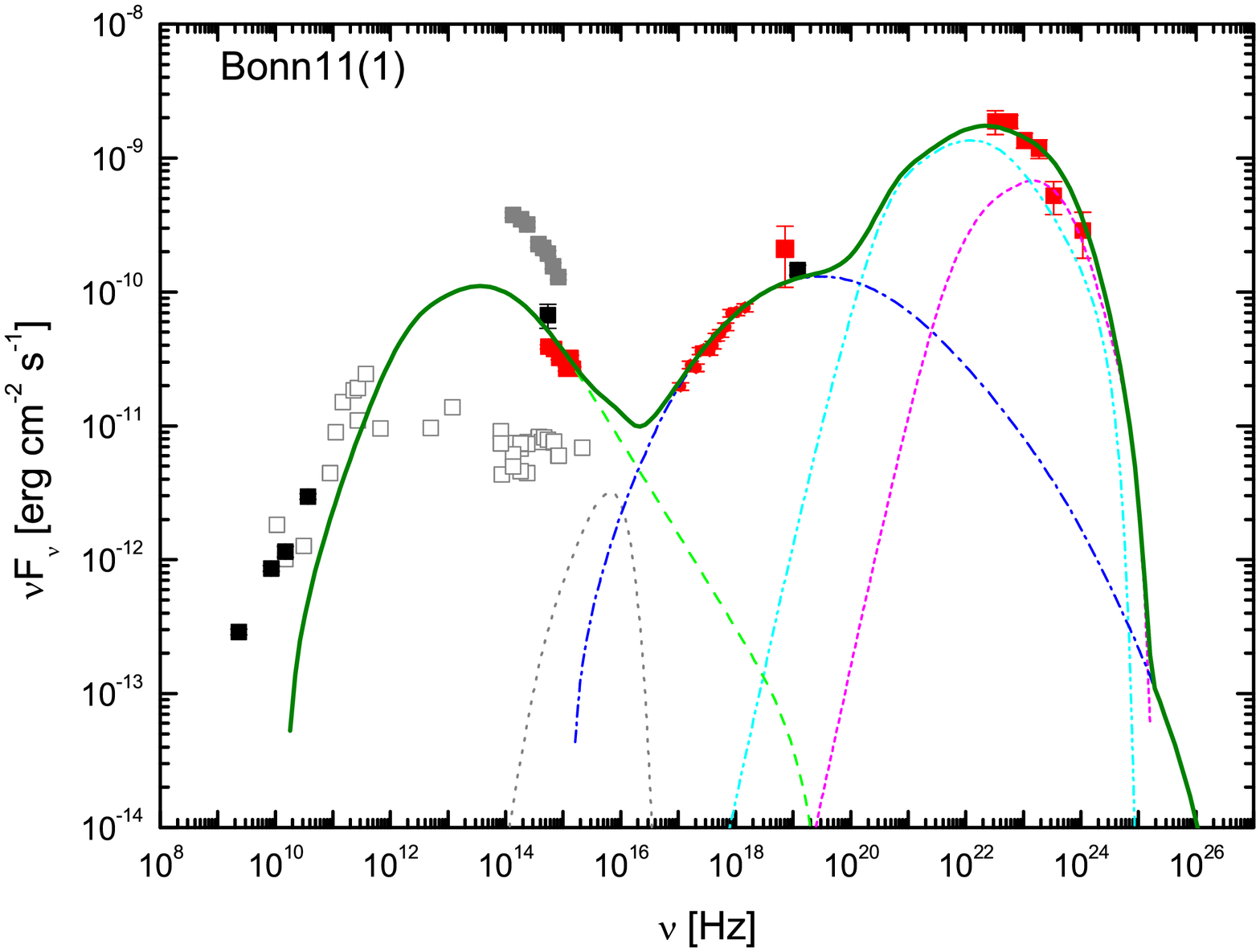}
\includegraphics[scale=0.28,angle=0]{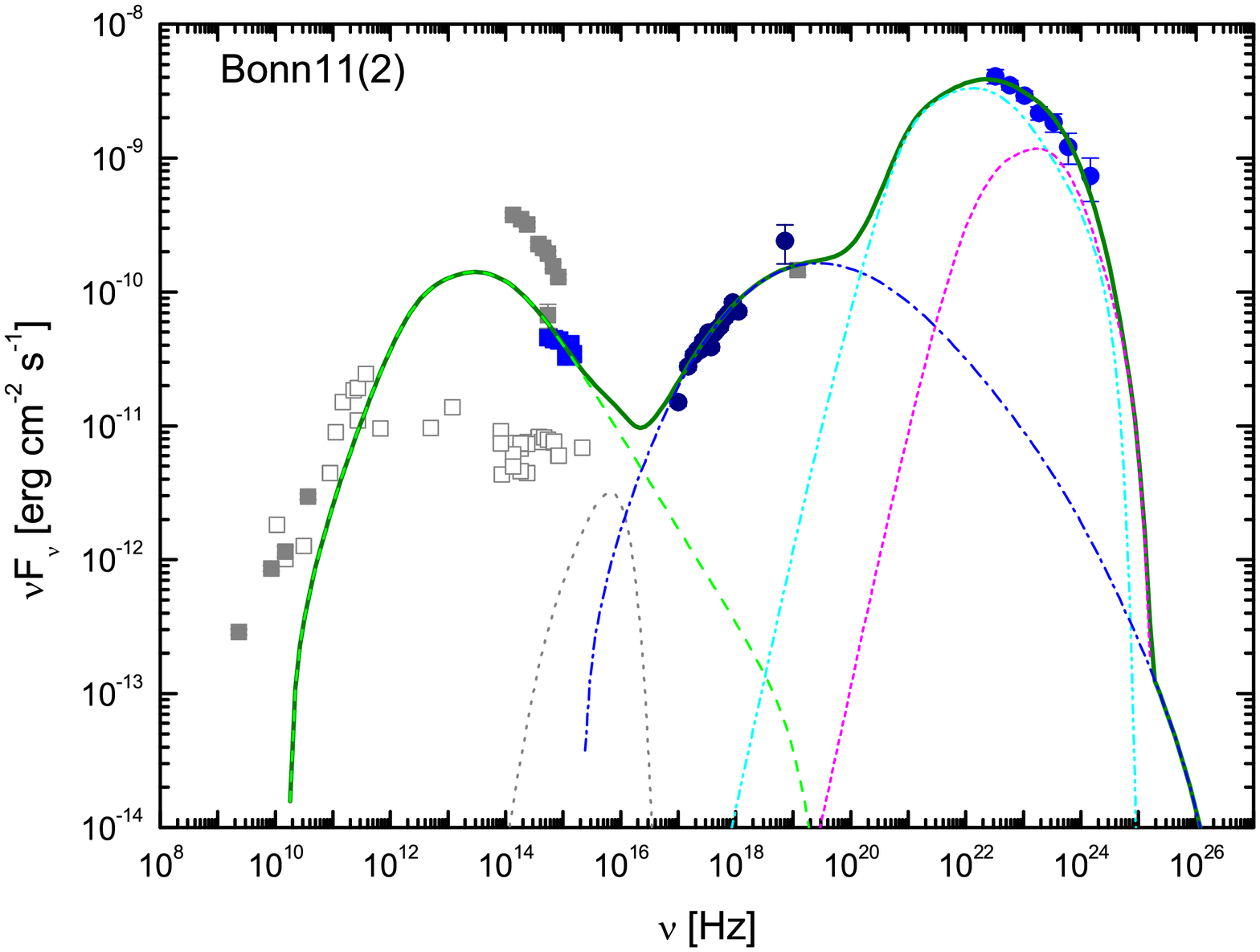}\\
\includegraphics[scale=0.28,angle=0]{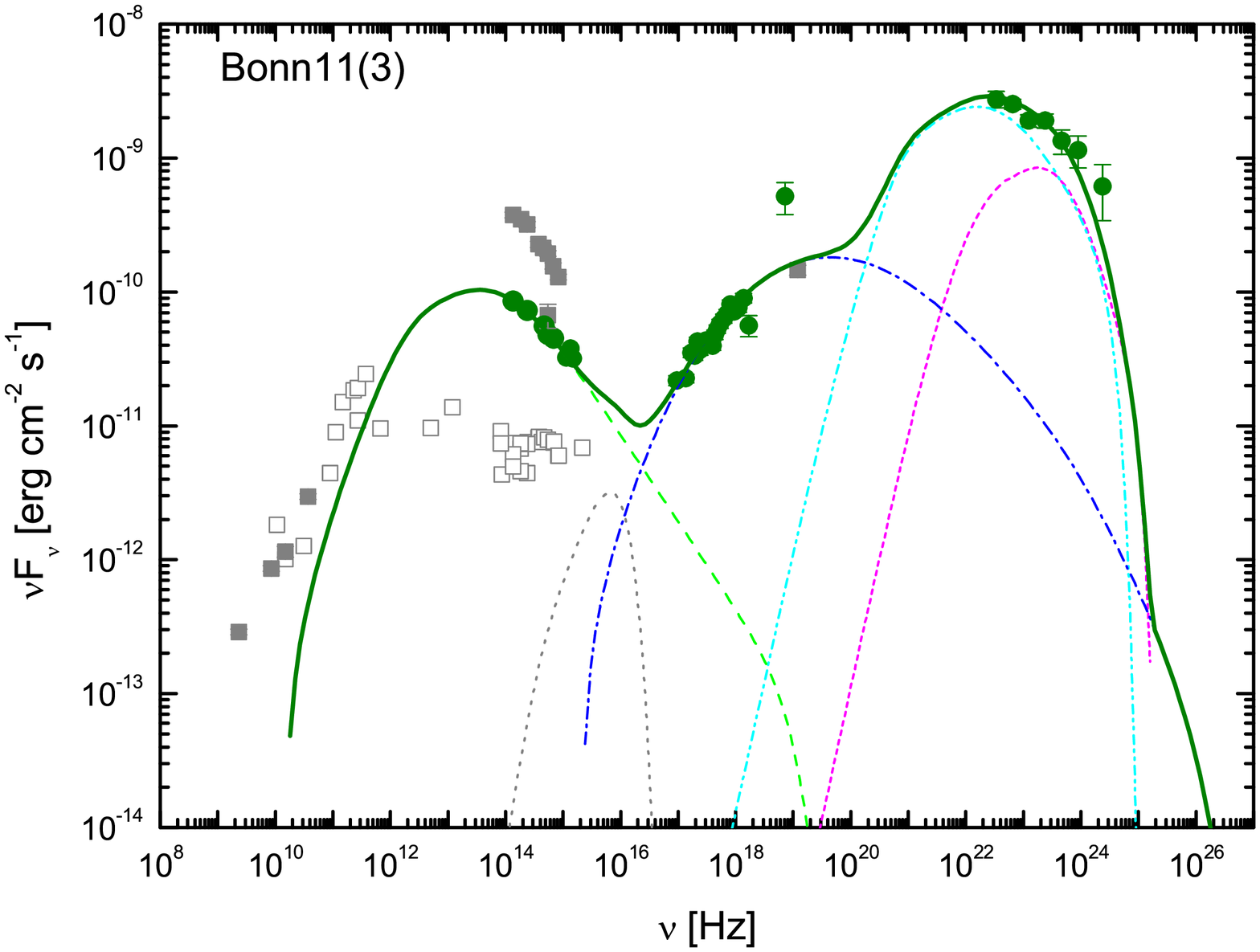}
\includegraphics[scale=0.28,angle=0]{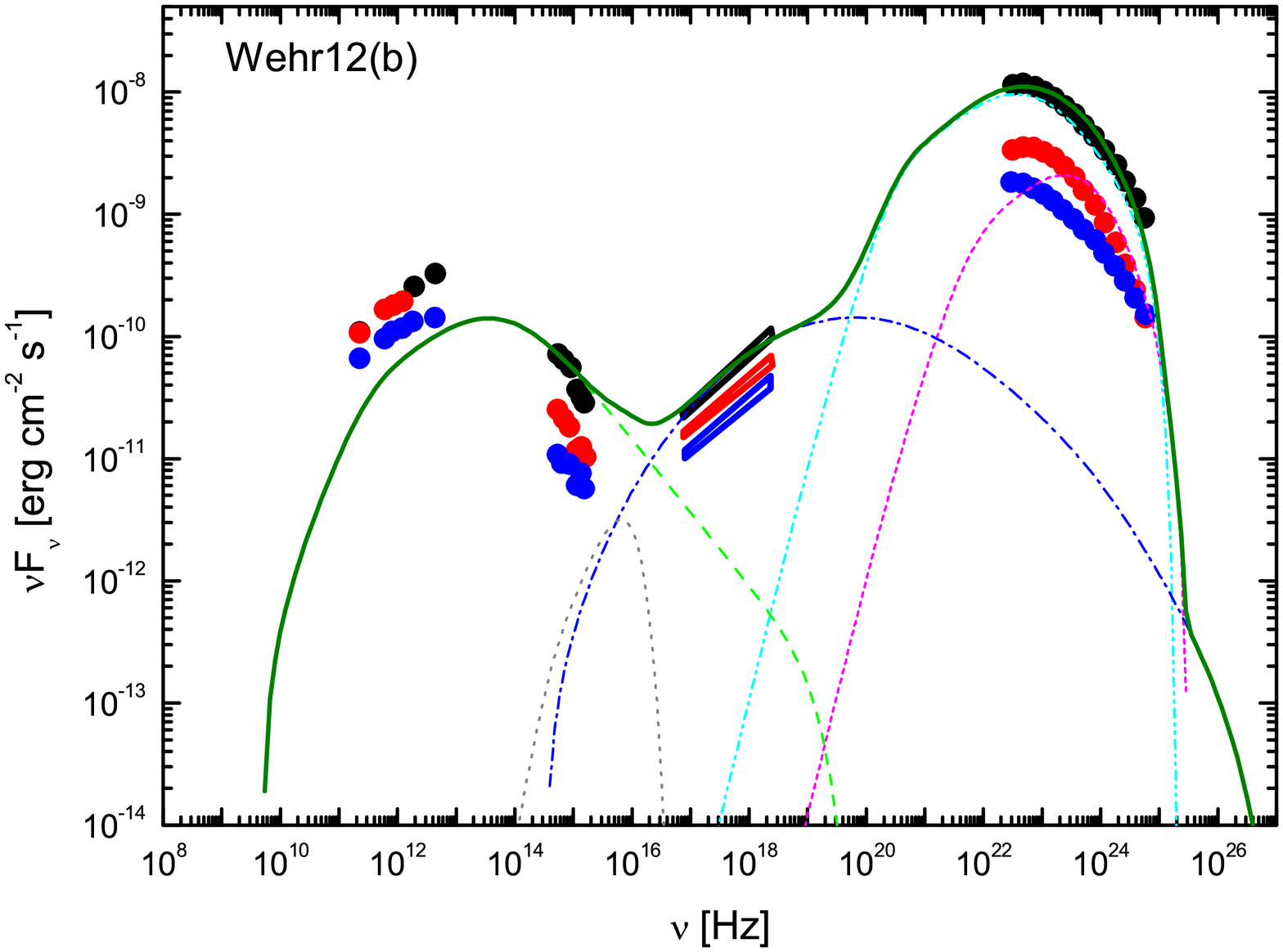}\\
\includegraphics[scale=0.28,angle=0]{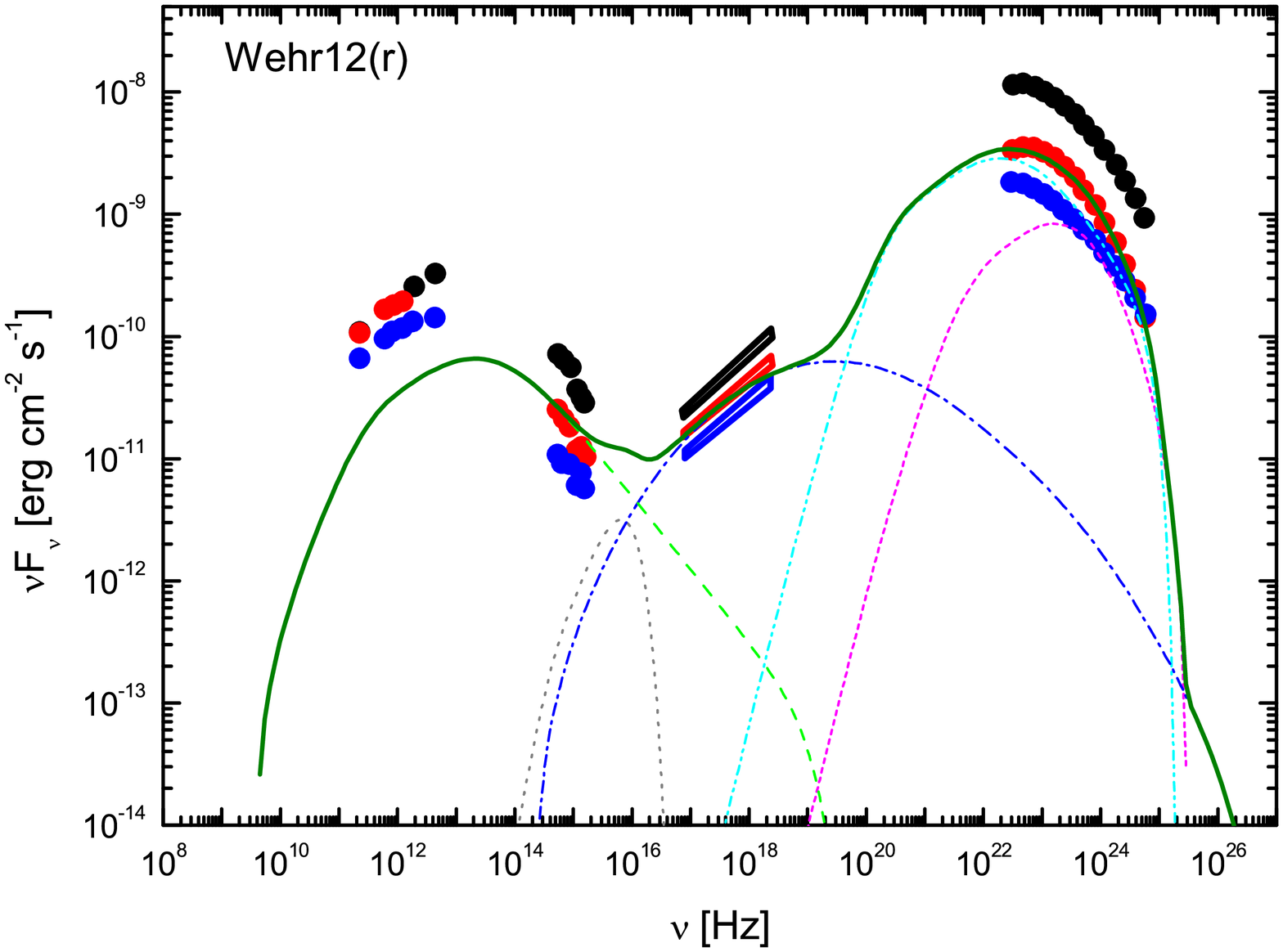}
\includegraphics[scale=0.28,angle=0]{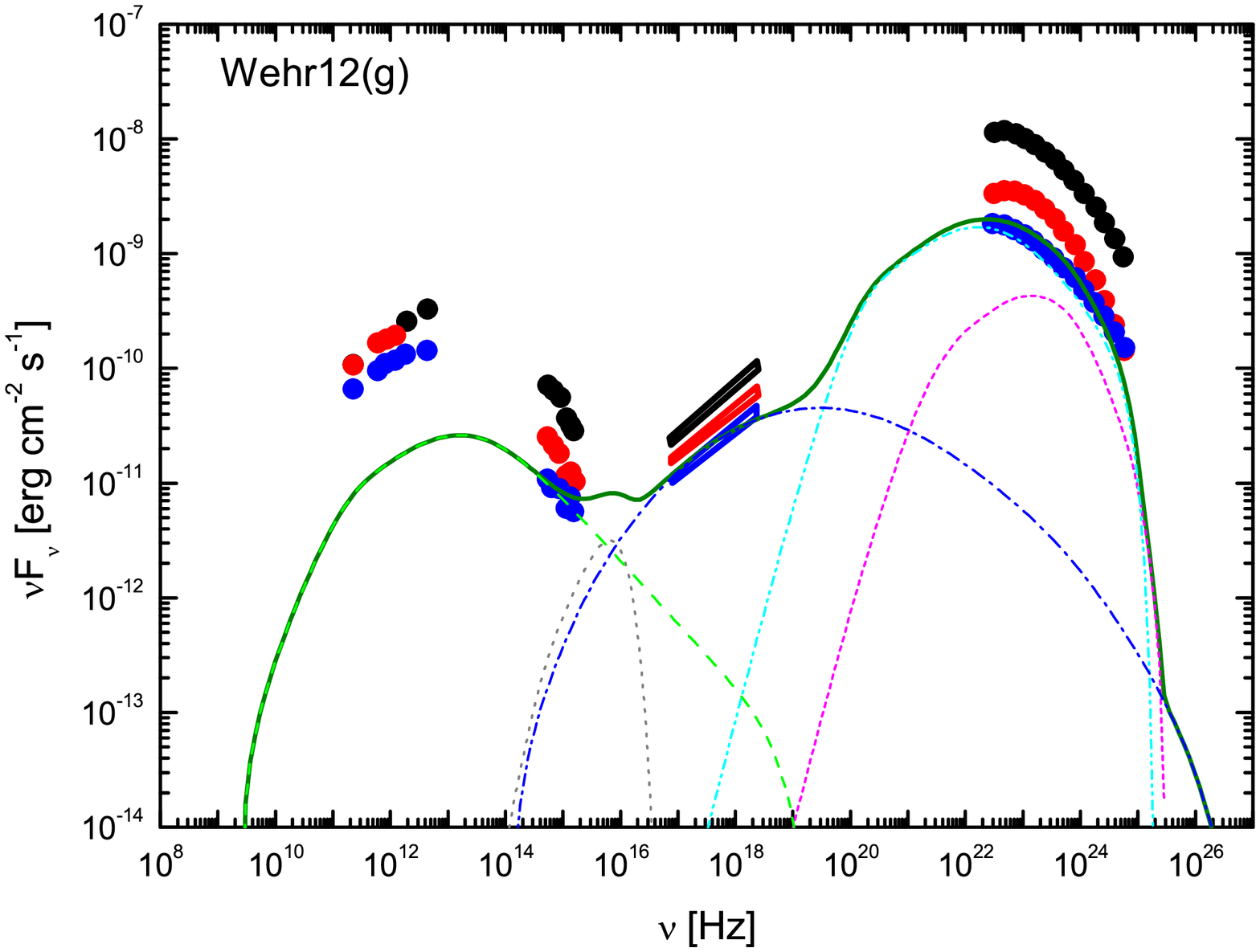}\\
\includegraphics[scale=0.28,angle=0]{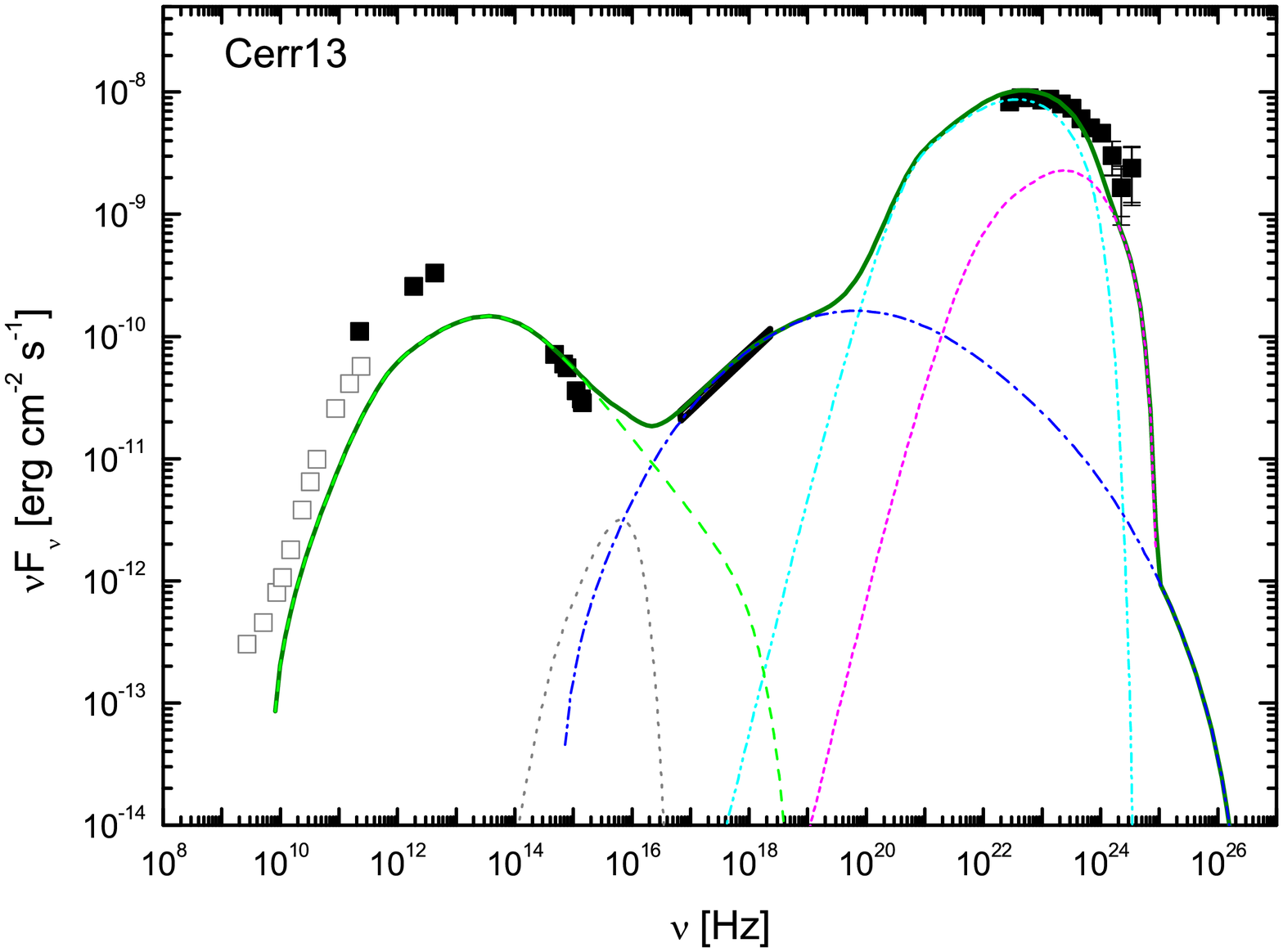}
\includegraphics[scale=0.28,angle=0]{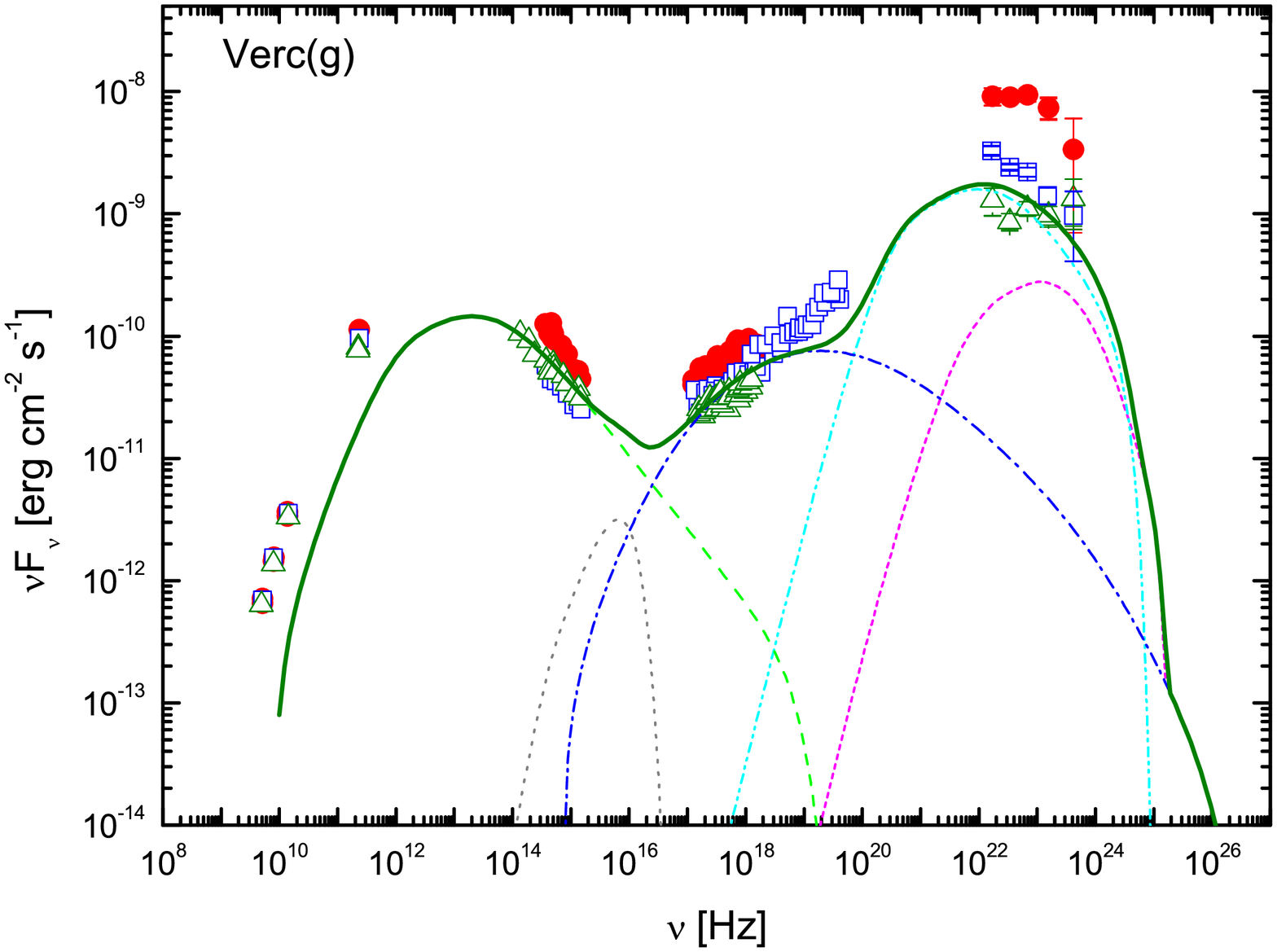}
\caption{Collected SEDs together with the fitting model, with the parameters listed in Table \ref{tab:tab-1}. The discrete points are the observed data. Separate spectral components (thin short dashed curves) are, from left to right, the synchrotron, the accretion disc, SSC, Compton scattering of the dust torus and the BLR radiations, respectively. The thick solid line represents the superposition of all the components.}
\label{fig:SED-3}
\end{figure*}

\begin{figure*}
\includegraphics[scale=0.28,angle=0]{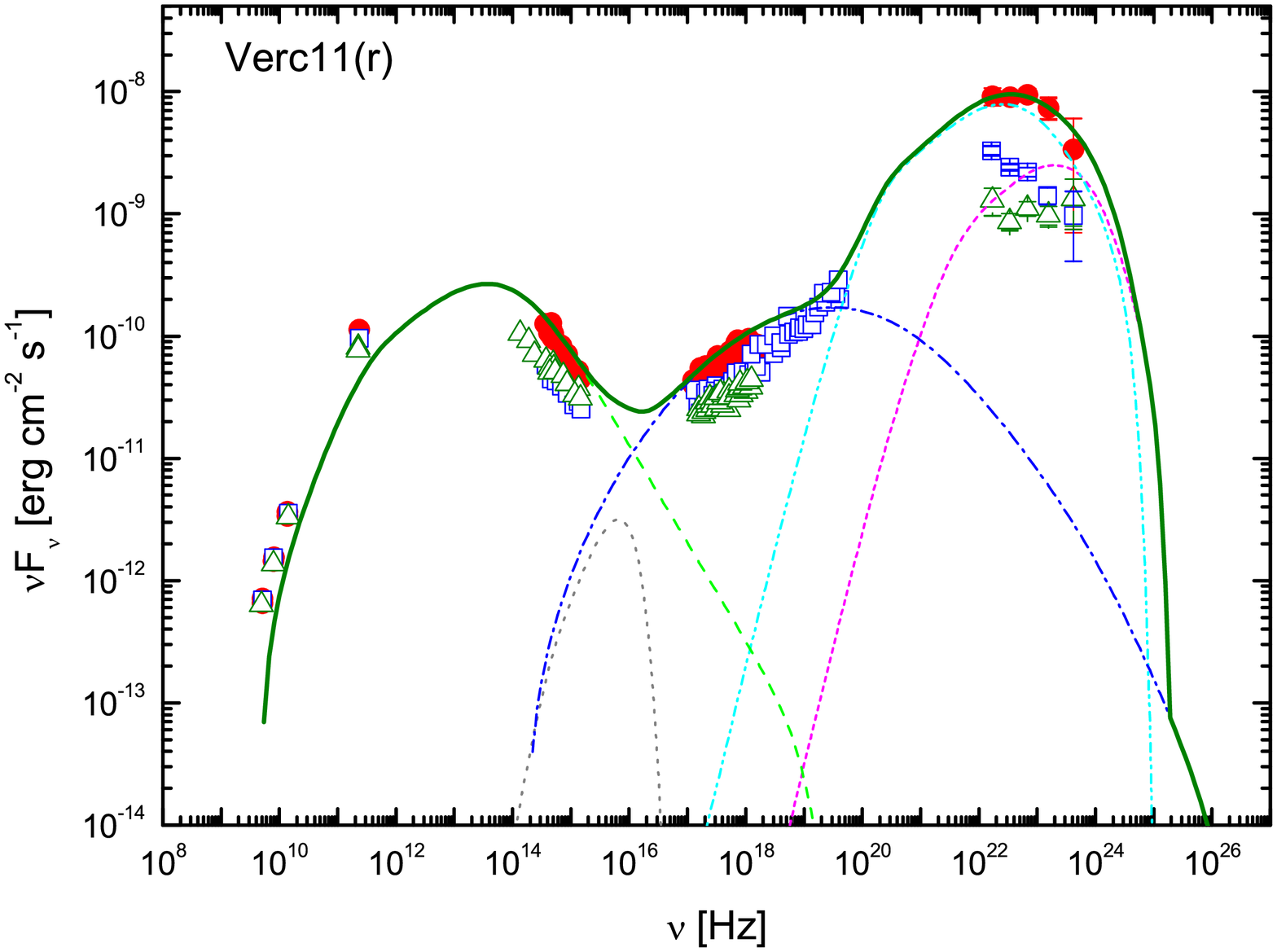}
\includegraphics[scale=0.28,angle=0]{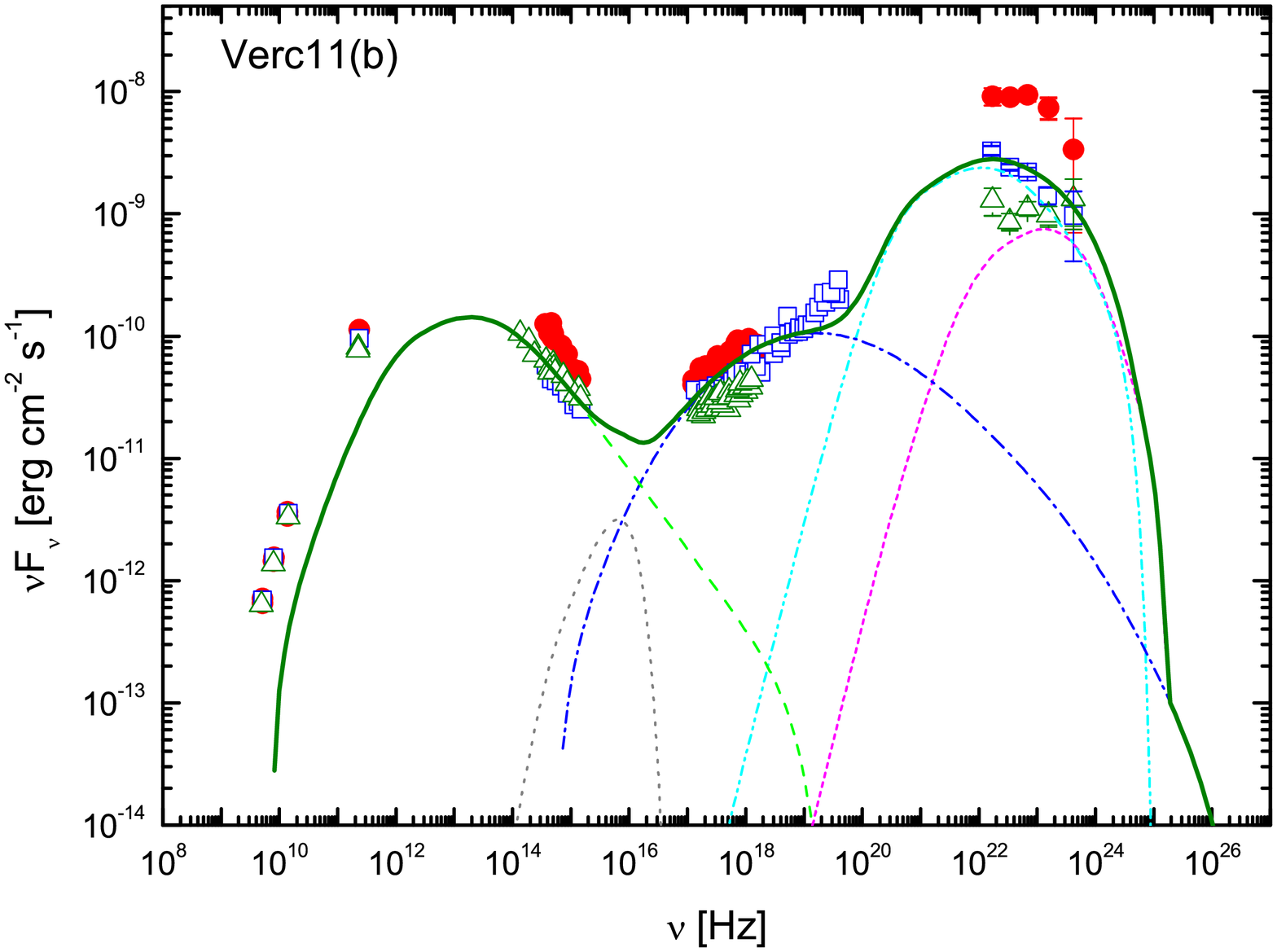}\\
\caption{Collected SEDs together with the fitting model, with the parameters listed in Table \ref{tab:tab-1}. The discrete points are the observed data. Separate spectral components (thin short dashed curves) are, from left to right, the synchrotron, the accretion disc, SSC, Compton scattering of the dust torus and the BLR radiations, respectively. The thick solid line represents the superposition of all the components.}
\label{fig:SED-4}
\end{figure*}

We also model the outburst with 3 hours short timescale variability (corresponding to $r_{\rm b}\simeq 6.7\times10^{-2}$\,pc). As stated in point (iv) discussed above, the $\gamma$-ray emitting region locates within the cavity of the BLR when $f_{\rm line}$ and $\tau_{\rm BLR}$ are assumed to be 0.05. Assuming that the emitting region has a Doppler factor $\delta_{\rm D}=25$, corresponding the bulk Lorentz factor to $\Gamma_{\rm j}\simeq13.5$, we find that the $\gamma$-ray photons with energy $E_{\gamma}=30$\,GeV can escape the BLR, as shown in Fig.\ref{fig:Colored-tau_E7.6-2}, supporting the flat structure of the BLR with $\alpha=45^{\circ}$.

In Table \ref{tab:tab-3}, we present typical peak frequencies, in which the synchrotron peaks appear at infrared to optical bands, and the SSC peaks are from X-rays to soft $\gamma$-ray bands. Because the Compton scattering of the BLR and the dust torus photons take place well in Thomson regime, our results show that the observed spectral break in the GeV spectrum reported by \cite{2009ApJ...699..817A} does not originate from K-N effect or relativistic electron distributions. From Fig.\ref{fig:SED-2}, we can see that the break is caused by the Compton scattering of dust torus photons over BLR photons, and the curved spectrum is the combination of Compton scattering of dust torus and BLR photons.

The emission region could be matter dominanted or magnetic-field dominanted, affecting the evolution of the jet \citep{1997MNRAS.288..833K,2007MNRAS.380...51K,2004ApJ...613..725S,2009ApJ...699.1274B}. Here, we explore this property of the emission regions during the outbursts. We define a equipartition parameter as $\eta_{\rm e}=u_{\rm e}^{\prime}/u_{\rm B}^{\prime}$. From the Table \ref{tab:tab-2}, it is clear that the $\eta_{\rm e}$ has the value from $\sim 0.2$ to $\sim 28$, in which 10 outbursts are dominated by magnetic energy, the rest 16 are dominated by particles, most outbursts take place in close equipartition. Moreover, we define a parameter $q_{\rm C}=L_{\rm EC}/L_{\rm syn}$ to measure the dominance of Compton emission, where $L_{\rm EC}$ is the luminosity of the SSC emission and EC emission by the BLR and the dust torus, $L_{\rm syn}$ is the synchrotron luminosity. From Table \ref{tab:tab-2}, we find that $q_{\rm C}$ has the value from $\sim0.1$ to $\sim 28$, implying that the emission region is Compton dominanted.

Assuming one proton per fifty electrons with $<\gamma_{\rm p}>\sim 1$ \citep{2000ApJ...534..109S}, we find that the total jet power $L_{\rm j}\simeq P_{\rm e}+p_{\rm p}+p_{\rm B}\simeq (0.9-18.5)\times 10^{46}$\,erg~s$^{-1}$. The Eddington luminosity $L_{\rm Edd}=4\pi GM_{\rm BH}m_{\rm p}c/\sigma_{\rm T}\simeq 6.3\times 10^{46}$\,erg~s$^{-1}$ for BH mass $M_{\rm BH}=5\times 10^{8}M_{\odot}$, indicating that the central engine works at high accretion rate or even super-Eddington rate during the outbursts.

The central BH mass of 3C 454.3 is still under debate, with $M_{\rm BH}=4.4\times 10^{9}$\,$M_{\odot}$ \citep{2001MNRAS.327.1111G}, or $M_{\rm BH}=5\times 10^{8}$\,$M_{\odot}$ \citep{2011MNRAS.410..368B}. In our model, we assume the inner and outer radius of the BLR  to be proportional to the BH mass. For $M_{\rm BH}=4.4\times 10^{9}$\,$M_{\odot}$, the diffuse energy density of the BLR is too low to produce the observed gamma-ray flux unless $f_{\rm BLR}$ is one orders of magnitude higher than the usual value. Our model favors the BH mass of 3C 454.3 to be $M_{\rm BH}=5\times 10^{8}$\,$M_{\odot}$.

\section{Conclusion and summary}
In this paper, we set up a leptonic model which includes SSC radiation and Compton scattering of external photons from a ``flat" BLR and the dust torus, and apply the model to 26 SEDs of FSRQ 3C 454.3 during the period from 2007 July to 2011 January. In the model, the accretion disc is approximated as a optically thick Shakura-Sunyaev one. The BLR has a flat structure described by an aperture angle $\alpha$ to reduce the absorption of $\gamma$-rays. The dust torus is simplified as a spherical shell.  Our main results are summarized as follows:

1. Both SSC and EC processes contribute to the high-energy components, but SSC emission only contributes to the low-energy part, while the high-energy part is dominated by EC processes. The emission regions are Compton dominanted with the parameter $q_{\rm C}=0.1 - 28$.

2. The distance of the $\gamma$-ray emission region from the BH, obtained from the observed variability timescale, are mostly within the BLR. At such distance, the electrons cooling are mainly dominated by soft photons either from the BLR or the dust torus.

3. The presence of a break in the GeV energy band would be mainly from Compton scattering of the BLR and the dust torus radiations.

4. Most $\gamma$-ray emission regions are in equipartition between magnetic and particle energy densities during the outbursts, in which $\eta_{\rm e}$ is from $\sim 0.2$ to $\sim 9$. Two outbursts show particle dominanted with $\eta_{\rm e}=15.1$ and 27.8.

5. The covering factor $f_{\rm BLR}$ of the BLR is inversely related to the covering factor $f_{\rm IR}$ of the dust torus, e.g., smaller $f_{\rm BLR}$ corresponds to larger $f_{\rm IR}$.

6. The aperture angle $\alpha$ of the BLR is about $45^{\circ}$ determined by the absorption of BLR photons to $\gamma$-rays.

7. The jet would be powered by accretion disc with high or even super-Eddington accretion rate.

8. The BH mass of 3C 454.3 would be $5\times 10^{8}$\,$M_{\odot}$.

\begin{center}
{Appendix} \\
{Derivation of the Doppler factor $\delta_{\rm D}$ and magnetic field $B$ in the Thomson regime }
\end{center}
In the SSC+EC model, according to \cite{1998ApJ...509..608T}, the Doppler factor, $\delta_{\rm D}$, and comoving magnetic field strength, $B$, could be estimated
from the peak $\nu F_{\nu}$ energy flux, $f_{\rm pk}^{\rm syn}$, $f_{\rm pk}^{\rm SSC}$ and $f_{\rm pk}^{\rm EC}$ of the synchrotron, SSC and EC components, respectively. Thus,
\begin{eqnarray}
\delta_{\rm D} &=& 12.6\biggr[\frac{f(\alpha_{1},\alpha_{2})}{3}\biggr]^{1/6}\biggr[\frac{10^{-4}\,\rm erg~cm^{-3}}{u_{\rm ext}}\biggr]^{1/6}\biggr[\frac{5\times 10^{16}\,\rm cm}{R_{\rm b}}\biggr]^{1/3}\biggr[\frac{d_{\rm L}}{10^{28}\, \rm cm}\biggr]^{1/3}
\nonumber\\
&\times&
\biggr[\frac{f_{\rm pk}^{\rm syn}}{10^{-10}\,\rm erg~cm^{-2}~s^{-1}}\biggr]^{1/6}\biggr[\frac{10^{-10}\,\rm erg~cm^{-2}~s^{-1}}{f_{\rm pk}^{\rm SSC}}\biggr]^{1/6}\biggr[\frac{f_{\rm pk}^{\rm EC}}{10^{-10}\,\rm erg~cm^{-2}~s^{-1}}\biggr]^{1/6},
\end{eqnarray}

\begin{eqnarray}
B &=& 0.63\biggr[\frac{f(\alpha_{1},\alpha_{2})}{3}\biggr]^{1/6}\biggr[\frac{u_{\rm ext}}{10^{-4}\,\rm erg~cm^{-3}}\biggr]^{1/3}\biggr[\frac{5\times 10^{16}\,\rm cm}{R_{\rm b}}\biggr]^{1/3}\biggr[\frac{d_{\rm L}}{10^{28}\,\rm cm}\biggr]^{1/3}
\nonumber\\
&\times&
\biggr[\frac{f_{\rm pk}^{\rm syn}}{10^{-10}\,\rm erg~cm^{-2}~s^{-1}}\biggr]^{2/3}
\biggr[\frac{10^{-10}\,\rm erg~cm^{-2}~s^{-1}}{f_{\rm pk}^{\rm SSC}}\biggr]^{1/6}\biggr[\frac{10^{-10}\,\rm erg~cm^{-2}~s^{-1}}{f_{\rm pk}^{\rm EC}}\biggr]^{1/3},
\end{eqnarray}
where $u_{\rm ext}$ is the energy density of the external radiation field.
$f(\alpha_{1},\alpha_{2})$ is
\begin{eqnarray}
f(\alpha_{1},\alpha_{2}) &=& \frac{1}{1-\alpha_{1}}+\frac{1}{\alpha_{2}-1}.
\end{eqnarray}

\section*{Acknowledgments}
We appreciate the anonymous referee for valuable comments that improved our paper significantly.
We acknowledge the financial supports from the National Natural Science Foundation of China 11133006, 11163006, 11173054, the Strategic Priority Research Program
¡°The Emergence of Cosmological Structures¡± of the Chinese
Academy of Sciences (XDB09000000), and the Policy Research Program of Chinese Academy of Sciences (KJCX2-YW-T24).

\begin{table*}
\caption{List of the parameters used to construct the theoretical SEDs. Notes: Column [1]: Magnetic field strength in units of G; Column [2]: The bulk Lorentz factor of the emitting region; Column [3]: The normalization of the electron distributions in units of cm$^{-3}$; Column [4]: The spectral index of the electron distribution above $\gamma_{\rm br}^{\prime}$. Column [5]-[6]: The minimum and the break energy of the electron energy distribution, respectively; Column [7]: the size of the emitting region. Column: [8]-[9] The covering factor of the BLR for lines and continuum, respectively. Column: [10] The covering factor of the dust torus. }
\label{tab:tab-1}
\begin{center}
\begin{tabular}{lcccccccccccccccc}
\hline
\hline
References &    B   & $\Gamma_{\rm j}$ & $n_{0}$  & $s_{2}$   &  $\gamma_{\rm min}^{\prime}$ &  $\gamma_{\rm br}^{\prime}$ & $R_{\rm b}$ &  $f_{\rm line}$ & $\tau_{\rm BLR}$ & $f_{\rm IR}$ \\
                 &  &     &           &     &   &($10^{2}$) &($10^{16}$) & & & & &\\
                 &[1]  &[2]    &[3]     &[4]     &[5]     &[6]     &[7]        &[8] &[9] &[10] \\
\hline
Verc09(12) &1.4  &15   &22       &4.1  &100       &8     &3.2    &0.1  &0.1  &0.25 \\

Verc09(13) &1.2  &13.5 &33       &4.2  &100       &8     &3.2    &0.1  &0.1  &0.25\\

Verc10(17) &1.4  &12.5 &82       &4.2  &150       &8.5   &2.12   &0.1  &0.1  &0.3\\

Verc10(18) &1.5  &13.5 &54       &4.5  &100       &8.5   &2.12   &0.1  &0.1  &0.3\\

Verc10(19) &1.6  &12.5 &39       &4.3  &120       &7.3   &2.12   &0.1  &0.1  &0.3 \\

Verc11(g)  &0.73 &15.5 &16       &4.2  &110       &8.3   &5.5    &0.1  &0.1  &0.25\\

Verc11(r)  &0.68 &25   &18       &4.6  &60        &11    &3.6    &0.1  &0.1  &0.3 \\

Verc11(b)  &0.65 &17.5 &29       &4.3  &105       &8.5   &4.2    &0.1  &0.1  &0.25\\

Donn09(r)  &1.7  &16.5 &97       &4.2  &110       &4.2   &2.1    &0.1  &0.1  &0.02\\

Donn09(g)  &2.2  &16   &85       &4.2  &120       &4     &2.2    &0.1  &0.1  &0.02\\

Donn09(b)  &1.9  &15.5 &93       &4.4  &110       &4     &2.2    &0.1  &0.1  &0.02\\

Pacc10(u-1)&0.3  &16.5 &28       &4.2  &240       &13    &5.3    &0.1  &0.1  &0.05\\

Pacc10(u-2)&0.35 &14.5 &23       &4.2  &240       &13    &5.3    &0.1  &0.1  &0.05\\

Pacc10(d-1)&0.36 &16   &23       &4.2  &220       &13    &5.3    &0.1  &0.1  &0.05\\

Pacc10(d-2)&0.4  &16   &14       &4.3  &200       &13    &5.3    &0.1  &0.1  &0.05\\

Abdo09     &0.46 &16   &3.4      &4.7  &150       &11    &7.69   &0.15  &0.15&0.03\\

Bonn11(05) &2.25 &13.5 &43       &4.5  &100       &8.5   &1.32   &0.04 &0.04 &0.65\\

Bonn11(6)  &1.15 &16.5 &106      &4.4  &110       &7.6   &1.56   &0.01 &0.01 &0.5\\

Bonn11(27) &1.05 &17   &100      &4.3  &130       &8.5   &1.74   &0.05 &0.05 &0.6\\

Bonn11(1)  &1.2  &18   &124      &4.4  &110       &8.8   &1.5    &0.05 &0.05 &0.6 \\

Bonn11(2)  &0.95 &20.5 &115      &4.4  &140       &8.5   &1.76   &0.05 &0.05 &0.6\\

Bonn11(3)  &0.85 &17.5 &119      &4.3  &140       &10    &1.77   &0.05 &0.05 &0.6\\

Wehr12(b)  &0.41 &22   &16       &4.2  &85        &13    &4.2    &0.1  &0.1  &0.3\\

Wehr12(r)  &0.40 &17.5 &21       &4.2  &85        &11    &4.2    &0.1  &0.1  &0.25\\

Wehr12(g)  &0.29 &13.5 &31       &4.1  &85        &12    &4.2    &0.1  &0.1  &0.3\\

Cerr13     &0.40 &22   &17.5     &4.2  &100       &13    &4.2    &0.1  &0.1  &0.25\\

\hline
\end{tabular}
\end{center}
References: Verc09(12), Verc09(12): shown in Fig.12 and Fig.13 by \cite{2009ApJ...690.1018V}; Verc10(17), Verc10(18), Verc10(19): shown by \cite{2010ApJ...712..405V};
Verc11(g), Verc11(r), Verc(b): shown with green, red and blue color by \cite{2011ApJ...736L..38V};
Donn09(r), Donn09(g), Donn09(b): shown with red, green and blue color by \cite{2009ApJ...707.1115D}; Pacc10(u-1), Pacc10(u-2), Pacc10(d-1), Pacc10(d-2): shown in the up panel and bottom panel by \cite{2010ApJ...716L.170P}; Abdo09: shown by \cite{2009ApJ...699..817A}; Bonn11(05), Bonn11(6),Bonn11(27), Bonn11(1), Bonn11(2), Bonn11(3): shown by \cite{2011MNRAS.410..368B}, in which ``05, 6, 27, 1, 2, 3" stand for date ``2005, Nov-6, Nov-27, Dec-1, Dec-2, Dec-3"; Wehr12(b), Wehr12(r), Wehr12(g): shown with black, red and blue color by \cite{2012ApJ...758...72W}; Cerr13: shown by \cite{2013ApJ...771L...4C}.
\label{tab:tab-1}
\end{table*}

\begin{table*}
\setlength{\tabcolsep}{0.02in}
\caption{The energy density of the electrons ($U_{\rm e}^{\prime}$) and magnetic field ($U_{\rm B}^{\prime}$); The jet power in the form of bulk motion of electron ($P_{\rm e}$), proton ($P_{\rm p}$) and Poynting flux ($P_{\rm B}$); Isotropic luminosity of synchrotron ($L_{\rm syn}$), SSC ($L_{\rm SSC}$); Compton-scattered dust torus ($L_{\rm EC}^{\rm torus}$) and BLR ($L_{\rm EC}^{\rm BLR}$) radiations; The total radiation luminosity ($P_{\rm r}$) and the total jet power ($P_{\rm j,tot}$), together with equipartition parameter ($\eta_{\rm e}$) and Compton dominance parameter ($q_{\rm C}$). Energy density is in units of erg~cm$^{-3}$. Power and luminosity are in units of erg~s$^{-1}$. In calculation to the power related to protons, assuming one proton per fifty electrons.}
\label{tab:tab-2}
\begin{center}
\begin{tabular}{lcccccccccccccccc}
\hline
\hline
References & $U_{\rm e,-2}^{\prime}$  & $U_{\rm B,-2}^{\prime}$  & $P_{\rm e,44}$  & $P_{\rm p,46}$  & $P_{\rm B,44}$ & $L_{\rm syn,48}$ &$L_{\rm SSC,48}$ & $L_{\rm EC,48}^{\rm IR}$ &$L_{\rm EC,48}^{\rm BLR}$ & $P_{\rm r,45}$  & $P_{\rm j,46}^{\rm tot}$ & $\eta_{\rm e}$ & $q_{\rm C}$  \\
\hline
Verc09(12) &3.25   &7.80   &7.06     &2.59   &16.93    &9.99  &5.55  &7.34  &3.39   &2.67     &2.83     &0.42    &1.63     \\

Verc09(13) &4.88   &5.73   &7.96     &2.92   &9.35     &6.53  &5.26  &5.23  &3.84   &2.58     &3.09     &0.85    &2.19     \\

Verc10(17) &4.88   &5.73   &8.58     &3.15   &10.08    &7.43  &5.99  &6.35  &3.07   &2.68     &3.34     &0.85    &2.07     \\

Verc10(18) &8.78   &8.95   &6.79     &2.49   &6.91     &5.74  &5.23  &3.84  &3.59   &2.16     &2.63     &0.98    &2.21     \\

Verc10(19) &4.59   &10.19  &3.04     &1.12   &6.74     &2.71  &1.33  &1.38  &1.31   &0.88     &1.21     &0.45    &1.48     \\

Verc11(g)  &2.37   &2.12   &16.24    &5.97   &14.52    &11.47 &8.24  &32.95 &4.66   &5.58     &6.27     &1.12    &4.00     \\

Verc11(r)  &5.47   &1.84   &41.78    &15.34  &14.06    &20.54 &18.49 &159   &43.14  &15.90    &15.90    &2.97    &0.11     \\

Verc11(b)  &4.59   &1.68   &23.35    &8.57   &8.56     &11.00 &11.20 &48.40 &12.32  &7.02     &8.89     &2.73    &6.54     \\

Donn09(r)  &4.29   &11.50  &5.33     &1.96   &14.28    &6.01  &2.15  &0.29  &4.05   &1.13     &2.15     &0.37    &1.08     \\

Donn09(g)  &3.61   &19.26  &4.22     &1.55   &22.49    &7.90  &2.42  &0.22  &2.98   &1.27     &1.82     &0.19    &0.71     \\

Donn09(b)  &4.10   &14.36  &4.49     &1.65   &15.74    &5.47  &1.74  &0.19  &2.84   &1.00     &1.85     &0.29    &0.87     \\

Pacc10(u-1)&5.44   &0.36   &39.23    &14.41  &2.58     &8.71  &24.66 &31.05 &19.34  &7.57     &14.82    &15.11   &8.62     \\

Pacc10(u-2)&4.47   &0.49   &24.87    &9.14   &2.71     &6.40  &14.88 &13.74 &8.91   &4.66     &9.41     &9.12    &5.86     \\

Pacc10(d-1)&4.76   &0.52   &32.23    &11.83  &3.49     &9.49  &22.35 &22.48 &14.26  &6.42     &12.19    &9.15    &6.23     \\

Pacc10(d-2)&3.07   &0.64   &20.82    &7.64   &4.31     &6.98  &9.83 &13.57  &8.86   &3.68     &7.90     &4.80    &3.19     \\

Abdo09     &0.76   &0.84   &11.39    &4.18   &12.64    &5.83  &2.36  &5.40  &3.55   &1.61     &4.42     &0.90    &1.94     \\

Bonn11(05) &6.99   &20.14  &2.09     &0.77   &6.03     &2.48  &1.12  &1.60  &0.66   &0.69     &0.85     &0.35    &1.36     \\

Bonn11(6)  &13.71  &5.26   &8.59     &3.14   &3.29     &4.08  &4.29  &10.66 &1.01   &1.81     &3.26     &2.61    &3.91     \\

Bonn11(27) &14.13  &4.39   &11.66    &4.28   &3.62     &6.27  &8.86  &24.28 &7.29   &4.08     &4.43     &3.22    &6.45     \\

Bonn11(1)  &20.15  &5.73   &13.84    &5.08   &3.94     &8.26  &13.32 &26.92 &10.68  &4.87     &5.26     &3.52    &6.16     \\

Bonn11(2)  &15.79  &3.59   &19.38    &7.12   &4.41     &10.20 &16.12 &63.08 &17.75  &7.85     &7.35     &4.40    &9.50     \\

Bonn11(3)  &21.10  &2.87   &19.08    &7.01   &2.60     &7.91  &18.91 &48.05 &13.18  &7.45     &7.22     &7.35    &10.13    \\

Wehr12(b)  &5.32   &0.67   &42.80    &15.72  &5.38     &11.85 &16.88 &207.8 &36.64  &19.07    &16.20    &7.94    &22.05    \\

Wehr12(r)  &5.50   &0.64   &27.99    &10.28  &3.24     &5.47  &7.26  &62.43 &14.70  &7.61     &10.59    &8.59    &15.43    \\

Wehr12(g)  &9.18   &0.33   &27.83    &10.22  &1.01     &2.27  &5.52  &38.66 &7.83   &6.36     &10.51    &27.81   &22.91    \\

Cerr13     &5.33   &0.64   &42.92    &15.76  &5.12     &12.16 &18.75 &170.7 &38.01  &16.72    &16.24    &8.33    &18.71    \\

\hline
\end{tabular}
\end{center}
References: See Table 1.
\label{tab:tab-2}
\end{table*}

\begin{table*}
\caption{Some typical frequencies. Column [1]: the peak of synchrotron radiation; Column [2]: the peak of SSC radiation; Column [3]: the peak of Compton-scattered BLR radiation; Column [4]: the peak of Compton-scattered dust torus radiation; Column [5]: The K-N critical position as Compton-scattered BLR radiation; Column [6]: The K-N critical position as Compton-scattered dust torus radiation.}
\label{tab:tab-3}
\begin{center}
\begin{tabular}{lcccccccccccc}
\hline
\hline
References & $\nu_{\rm syn,13}$  & $\nu_{\rm SSC,19}$  & $\nu_{\rm EC}^{\rm BLR}$ & $\nu_{\rm EC}^{\rm IR}$ & $\nu_{\rm KN}^{\rm BLR}$ & $\nu_{\rm KN}^{\rm IR}$\\
&          &      &   (GeV)                  &      (GeV)              &      (GeV)               &    (TeV)  \\
&   [1]               &    [2]              &   [3]                    &      [4]                &      [5]                 &    [6]             \\
\hline
 Verc09(12) &4.87    &4.14     &1.91      &$2.32\times10^{-2}$     &14.68     &1.17\\

 Verc09(13) &3.82    &3.25     &1.58      &$1.92\times10^{-2}$     &14.93     &1.19\\

 Verc10(17) &4.70    &4.52     &1.54      &$1.87\times10^{-2}$     &15.08     &1.21\\

 Verc10(18) &5.39    &5.18     &1.78      &$2.16\times10^{-2}$     &14.93     &1.19\\

 Verc10(19) &3.97    &2.81     &1.14      &$1.38\times10^{-2}$     &15.08     &1.21\\

 Verc11(g)  &2.81    &2.57     &2.19      &$2.66\times10^{-2}$     &14.59     &1.17\\

 Verc11(r)  &6.43    &10.3     &8.67      &$0.11             $     &12.66     &1.01\\

 Verc11(b)  &2.88    &2.77     &2.85      &$3.46\times10^{-2}$     &14.22     &1.14\\

 Donn09(r)  &1.76    &0.41     &0.63      &$7.61\times10^{-3}$     &14.41     &1.15\\

 Donn09(g)  &2.01    &0.43     &0.54      &$6.53\times10^{-3}$     &14.50     &1.16\\

 Donn09(b)  &1.70    &0.36     &0.51      &$6.17\times10^{-3}$     &14.59     &1.17\\

 Pacc10(u-1)&2.97    &6.68     &6.00      &$7.29\times10^{-2}$     &14.41     &1.15\\

 Pacc10(u-2)&3.12    &7.02     &4.75      &$5.77\times10^{-2}$     &14.76     &1.18\\

 Pacc10(d-1)&3.48    &7.83     &5.68      &$6.90\times10^{-2}$     &14.50     &1.16\\

 Pacc10(d-2)&3.87    &8.69     &5.68      &$6.90\times10^{-2}$     &14.50     &1.16\\

 Abdo09     &3.18    &5.13     &4.07      &$4.94\times10^{-2}$     &14.50     &1.16\\

 Bonn11(05) &8.08    &7.76     &1.78      &$2.16\times10^{-2}$     &14.93     &1.19\\

 Bonn11(6)  &3.89    &2.99     &2.05      &$2.49\times10^{-2}$     &14.41     &1.15\\

 Bonn11(27) &4.55    &4.38     &2.71      &$3.29\times10^{-2}$     &14.31     &1.15\\

 Bonn11(1)  &5.83    &6.00     &3.21      &$3.90\times10^{-2}$     &14.12     &1.13\\

 Bonn11(2)  &4.73    &4.54     &3.75      &$4.55\times10^{-2}$     &13.62     &1.09\\

 Bonn11(3)  &5.22    &6.94     &3.94      &$4.79\times10^{-2}$     &14.22     &1.14\\

 Wehr12(b)  &5.00    &11.25    &9.86      &$0.12$                  &13.31     &1.06\\

 Wehr12(r)  &2.97    &4.78     &4.77      &$5.80\times10^{-2}$     &14.22     &1.14\\

 Wehr12(g)  &2.08    &3.98     &3.55      &$4.31\times10^{-2}$     &14.93     &1.19\\

 Cerr13     &4.88    &10.97    &9.86      &0.12                    &13.31     &1.06\\

\hline
\end{tabular}
\end{center}
References: See Table 1.
\label{tab:tab-3}
\end{table*}

\end{document}